\documentclass[a4paper, 11pt]{article}
\usepackage{amsfonts,slashed}
\usepackage{float}
\usepackage{cite}
\usepackage{latexsym}
\usepackage{amsfonts}
\usepackage{amsmath}
\usepackage{epsfig}
\usepackage{latexsym,amssymb}
\usepackage{amsmath,amssymb,amsthm}
\usepackage{hyperref}
\usepackage{graphicx,tikz}
\usepackage{multirow}
\usepackage{textcomp}
\usepackage{xcolor}
\usepackage{mathtools}
\usepackage{stmaryrd}
\usepackage{enumerate}
\usepackage[shortlabels]{enumitem}

\usepackage{ifpdf}
\ifx\pdfoutput\undefined
\pdffalse
\usepackage{cite}
\else
\pdfoutput=1
\pdftrue
\fi
\DeclareFontFamily{OMS}{rsfs}{\skewchar\font'60}
\DeclareFontShape{OMS}{rsfs}{m}{n}{<-5>rsfs5 <5-7>rsfs7 <7->rsfs10 }{}
\DeclareSymbolFont{rsfs}{OMS}{rsfs}{m}{n}
\DeclareSymbolFontAlphabet{\Scr}{rsfs}

\numberwithin{equation}{section}
\def\be{\begin{equation}}
	\def\ee{\end{equation}}
\def\ba{\begin{array}}
	\def\ea{\end{array}}

\newcommand{\bea}{\begin{eqnarray}}
	\newcommand{\eea}{\end{eqnarray}}

\newcommand{\ff}[1]{\mathfrak{#1}}
\newcommand{\fA}{\ff{A}}

\newcommand{\FF}{{\cal F}}
\newcommand{\FG}{{\cal G}}
\newcommand{\ov}[1]{\overline{#1}}

\newcommand{\ON}[1]{\mathrm{O}( #1 )}
\newcommand{\SU}[1]{\mathrm{SU}( #1 )}
\newcommand{\SL}[1]{\mathrm{SL}( #1 )}

\newcommand{\SO}[1]{\mathrm{SO}( #1 )}

\newcommand{\U}[1]{\mathrm{U}(#1)}
\newcommand{\Spin}[1]{\mathrm{Spin}(#1)}
\newcommand{\EE}{E_{8(8)}}
\newcommand{\USp}[1]{\mathrm{USp}(#1)}

\newcommand{\Gt}{\mathrm{G}_{2(2)}}
\newcommand{\Ff}{\mathrm{F}_{4(4)}}
\newcommand{\En}[1]{E_{#1(#1)}}
\newcommand{\Gst}{G_S}
\newcommand{\Msc}{{\cal M}_{\rm scalar}}
\newcommand{\SUc}[2]{\mathrm{S}(\U{#1}\times\U{#2})}

\newcommand{\Com}[2]{\mathrm{Com}_{#1}(#2)}
\newcommand{\SLA}{\SL{8}_{\rm IIA}}
\newcommand{\SLB}{\SL{8}_{\rm IIB}}

\newcommand{\su}[1]{\mathfrak{su}(#1)}
\newcommand{\so}[1]{\mathfrak{so}(#1)}

\newcommand{\mbf}[1]{\mathbf{#1}}
\newcommand{\mbfb}[1]{\mathbf{\left(#1\right)}}
\newcommand{\obf}[1]{\mbf{\overline{\mathbf{#1}}}}
\newcommand{\trep}[1]{(\mbf{#1})}
\newcommand{\PR}[1]{(\mathbb{P}_{#1})}

\newcommand{\Leib}[2]{\llbracket #1,#2 \rrbracket}

\newcommand{\+}{\oplus}

\newcommand{\gL}{\mathcal{L}}
\newcommand{\gM}{\mathcal{M}}

\DeclareMathOperator{\tr}{Tr}
\newcommand{\EM}[1]{}
\newcommand{\MG}[1]{}
\newcommand{\GaGr}{$\mathrm{ISO}(3) \times \U{1}^4$}
\newcommand{\CG}{\Com{G}{\EE}}
\newcommand{\cJ}{{\cal J}}

\begin{document}
	
\begin{titlepage}
	\vfill
	\begin{flushright}
		HU-EP-22/23
	\end{flushright}
	
	\vfill

	\begin{center}
		\baselineskip=16pt
		{\Large \bf Consistent truncations to 3-dimensional supergravity}
		\vskip 2cm
		{\large \bf Michele Galli$^a$\footnote{\tt michele.galli@physik.hu-berlin.de},  Emanuel Malek$^a$\footnote{\tt emanuel.malek@physik.hu-berlin.de}}
		\vskip .6cm
		{\it $^a$ Institut f\"ur Physik, Humboldt-Universit\"at zu Berlin, \\			IRIS Geb\"aude, Zum Gro\ss en Windkanal 2, 12489 Berlin, Germany \\ \ \\}
		\vskip 1cm
	\end{center}
	\vfill
		
	\begin{abstract}
		We show how to construct consistent truncations of 10-/11-dimensional supergravity to 3-dimensional gauged supergravity, preserving various amounts of supersymmetry. We show, that as in higher dimensions, consistent truncations can be defined in terms of generalised $G$-structures in Exceptional Field Theory, with $G \subset \En{8}$ for the 3-dimensional case. Differently from higher dimensions, the generalised Lie derivative of $\En{8}$ Exceptional Field Theory requires a set of ``covariantly constrained'' fields to be well-defined, and we show how these can be constructed from the $G$-structure itself. We prove several general features of consistent truncations, allowing us to rule out a higher-dimensional origin of many 3-dimensional gauged supergravities. In particular, we show that the compact part of the gauge group can be at most $\SO{9}$ and that there are no consistent truncations on a 7-or 8-dimensional product of spheres such that the full isometry group of the spheres is gauged. Moreover, we classify which matter-coupled ${\cal N} \geq 4$ gauged supergravities can arise from consistent truncations. Finally, we give several examples of consistent truncations to three dimensions. These include the truncations of IIA and IIB supergravity on $S^7$ leading to two different ${\cal N}=16$ gauged supergravites, as well as more general IIA/IIB truncations on $H^{p,7-p}$. We also show how to construct consistent truncations on compactifications of IIB supergravity on $S^5$ fibred over a Riemann surface. These result in 3-dimensional ${\cal N}=4$ gauged supergravities with scalar manifold ${\cal M} = \frac{\SO{6,4}}{\SO{6} \times \SO{4}} \times \frac{\SU{2,1}}{\SUc{2}{1}}$ with a \GaGr{} gauging and for hyperboloidal Riemann surfaces contain ${\cal N}=(2,2)$ AdS$_3$ vacua.
		\end{abstract}
	\vskip 0cm
		
	\vfill
		
	\setcounter{footnote}{0} 
		
\end{titlepage}
	
\tableofcontents
	
\newpage
	
\section{Introduction}
Finding consistent Kaluza-Klein (KK) truncations of $10$- and $11$-dimensional supergravity to lower dimensions is, in general, extremely complicated. One needs to truncate the infinite tower of KK fluctuations on the compactification in such a way that a solution of the lower-dimensional, highly non-linear, equations of motion also solves the higher-dimensional ones \cite{Duff:1984hn,Cvetic:2000dm}. In practice, this means finding an inspired truncation ansatz such that, upon substitution into the equations of motion, all the dependence on the compactification coordinates factorises.

However, in recent years, a systematic approach to consistent truncations came from the Exceptional Field Theory (ExFT) and Exceptional Generalised Geometry (EGG) approaches \cite{Lee:2014mla,Hohm:2014qga,Baguet:2015sma,Malek:2016bpu,Malek:2017njj,Malek:2019ucd,Cassani:2019vcl,Cassani:2020cod}.\footnote{For the purposes of this paper, the ExFT and EGG approaches coincide, since we will be solving the ``section condition'' of ExFT globally. The only difference between the two formalisms arises, if one allows for different solutions of the section condition in disparate patches ExFT, meant to model non-geometric backgrounds, or if one allows for violations of the section condition. We will not consider these possibilities here and instead focus on 10-/11-dimensional supergravity.} These formalisms unify the metric and flux degrees of freedom of 10-/11-dimensional supergravity, leading to an otherwise hidden $\En{d}$ symmetry group of the theory. Consistent truncations can now be constructed using group theory and the language of ``$G$-structures'': a consistent truncation can be defined on a compactification that admits a reduced generalised $G \subset \En{d}$-structure, whose intrinsic torsion only contains constant singlets under $G$ \cite{Malek:2017njj,Cassani:2019vcl}. The truncation ansatz can then easily be constructed by expanding the ExFT fields in terms of the generalised $G$-structure, while the intrinsic torsion defines the embedding tensor of the lower-dimensional gauged supergravity.

For example, consistent truncations to maximal gauged supergravity arise from generalised identity structures. This means that one has a globally well-defined generalised frame in ExFT. The intrinsic torsion condition in this case requires the generalised frame to close into an algebra under the generalised Lie derivative. This leads to the generalised Scherk-Schwarz ansatz \cite{Aldazabal:2011nj,Geissbuhler:2011mx,Grana:2012rr,Berman:2012uy,Geissbuhler:2013uka,Berman:2013cli,Hohm:2014qga,Lee:2014mla}. Similarly, consistent truncations to lower-dimensional gauged supergravities with less supersymmetry require a generalised $G$-structure, that stabilises ${\cal N}$ spinors in ExFT, with ${\cal N}$ determining the amount of supersymmetry of the lower dimensional supergravity. For example, consistent truncations preserving half of the supersymmetries in $11-d$ dimensions, $ d \geq 7$, are described by a generalised $\Spin{d-1-n}$ structure, with $n$ labelling the number of vector multiplets \cite{Malek:2016bpu,Malek:2017njj}.

This new viewpoint both captures existing consistent truncations, such as of 11-dimensional supergravity on $S^7$ \cite{deWit:1986oxb}, and also yields many new ones \cite{Hohm:2014qga,Lee:2014mla,Malek:2015hma,Baguet:2015sma,Baguet:2015iou,Lee:2015xga,Malek:2016bpu,Ciceri:2016dmd,Cassani:2016ncu,Inverso:2016eet,Malek:2017cle,Malek:2017njj,Malek:2018zcz,Malek:2019ucd,Cassani:2019vcl,Malek:2020jsa,Cassani:2020cod}. It has also allowed us to rule out the higher-dimensional origin of large classes of gauged supergravities. For example, there are universal upper bounds on the number of matter multiplets in gauged supergravities that arise from consistent truncations \cite{Malek:2017njj,Cassani:2019vcl,Josse:2021put}.

The situation is quite different for consistent truncations to three dimensions. On the one hand, the duality between scalars and vectors leads to a particularly large number of possible three-dimensional gauged supergravities. These can have enormous gauge groups, including $\EE, \,\SO{8}\times\SO{8},\, F_{4(4)}\times\Gt,\, \SO{8,\mathbb{C}}$ \cite{Nicolai:2000sc,Nicolai:2001sv,Fischbacher:2003yw,deWit:2003fgi}, which, unlike in higher-dimensional gauged supergravities, cannot arise purely from the isometries of the compactification. On the other hand, the $\EE$ ExFT approach is less clear in three dimensions, since the algebra of the generalised Lie derivative does not close even upon solving the section condition. Instead, one must include an additional gauge parameter, the ``covariantly-constrained'' field $\Sigma$, which compensates the anomaly of the generalised Lie bracket. The need for covariantly-constrained fields and lack of closure of the generalised Lie derivative make it less clear how to define $G$-structures and their intrinsic torsion in this setup. Nonetheless, there clearly should be a generalisations of these concepts to the three-dimensional case. Indeed, \cite{Hohm:2017wtr} showed that a generalised Scherk-Schwarz ansatz of $\ON{8,8}$ double field theory can be used to construct consistent truncations of half-maximal 10-dimensional supergravity to three dimensions.

In this paper, we will show how to use the $\EE$ ExFT machinery to systematically construct consistent truncations of maximally supersymmetric 10/11-dimensional supergravities to three dimensions, preserving different amounts of supersymmetry. We will derive general results about which theories can be obtained by consistent truncations, placing restrictions on both the gaugings and matter contents. We classify the possible matter content of all $\mathcal{N}> 4$ theories with uplifts to 10-/11-dimensional supergravity and list their associated structure groups. We also give a partial classification for $\mathcal{N}=4$, including the structure groups and matter contents corresponding to all single quaternionic Kähler scalar manifolds and for some examples of products of quaternionic Kähler manifolds.

We will construct several new examples of consistent truncations. These include truncation of IIA/IIB supergravity preserving maximal supersymmetry with $\SO{8} \ltimes T^{28}$ gauging, and different real forms thereof. We will also give an example of a consistent truncation of type IIB on $S^5$ twisted over a Riemann surface, leading to a $\mathcal{N}=4$ gauged supergravity with scalar manifold $\frac{\SO{6,4}}{\SO{6} \times \SO{4}} \times \frac{\SU{2,1}}{\SUc{2}{1}}$ and $\mathrm{ISO}(3)\times \U{1}^4$ gauging. When the Riemann surface is a hyperboloid, this theory contain ${\cal N} = (2,2)$ AdS$_3$ vacua that arise from the near-horizon limit of D5-branes compactified on a Riemann surface.

This paper is structured as follows. We begin with a review of the salient features of $\En{8}$ ExFT in section \ref{s:E8ExFT}, before studying the generalised Lie derivative and its generalised Killing vector fields in \ref{s:GenKill}, since these play an important role in the gauging obtained from consistent truncations. In section \ref{s:GSS}, we then describe how to construct consistent truncations to 3-dimensional maximal gauged supergravity, and show that the form of the generalised Lie derivative places strong constraints on the possible gaugings. In particular, the compact part of the gauging can be at most $\SO{9}$, and we also prove that we cannot construct a consistent truncation on the 7- or 8-dimensional product of spheres, such that the full isometry group is gauged, unlike in the 4-dimensional case. We then show in section \ref{s:S7} how to construct consistent truncations from twist matrices valued in one of two $\SL{8} \subset \En{8}$ subgroups, corresponding to IIA/IIB truncations. This allows us to construct the consistent truncation of IIA/IIB on $S^7$ to two different three-dimensional ${\cal N}=16$ $\SO{8} \ltimes T^{28}$ gauged SUGRAs, as well as IIA/IIB truncations on $H^{p,7-p}$. We also discuss how the IIB truncation on $S^7$ is related to the IIA one by an outer automorphism of $\SO{8,8}$ in subsection \ref{s:Outer}.

Then, in section \ref{s:LessSUSY}, we show how to describe consistent truncations to a gauged supergravity with less than maximal supersymmetry. In section \ref{s:ClassLessSUSY}, we classify all possible matter contents that can arise for ${\cal N} > 4$ gauged SUGRA and give the relevant structure groups. For ${\cal N}=4$ gauged SURAs whose scalar manifold is a single quaternionic-K\"{a}hler manifold, we again fully classify which can arise from a consistent truncation based on the structure group, while we only partially classify those whose scalar manifolds are a product of two quaternionic-K\"{a}hler manifolds. Finally, in section \ref{s:Riemann}, we use this formalism to construct the consistent truncation of IIB SUGRA on $S^5$ fibred over a Riemann surface. The resulting theory has ${\cal N} = 4$ supersymmetry with scalar manifold $\frac{\SO{6,4}}{\SO{6} \times \SO{4}} \times \frac{\SU{2,1}}{\SUc{2}{1}}$ and gauging \GaGr. We conclude with a discussion and outlook in section \ref{s:Conclusions}.

\section{$\EE$ Exceptional Field Theory}\label{s:E8ExFT}

\subsection{Review of $\EE$ gauge structure and Lagrangian}\label{s:Rev}
	
The $\EE$ ExFT \cite{Hohm:2014fxa} consists of the following bosonic fields
\begin{equation} \label{eq:BosField}
	\left\{ g_{\mu\nu},\,\gM_{MN},\, A_\mu{}^M,\, B_{\mu\,M} \right\} \,, \qquad \mu,\nu = 0, \ldots, 2\,, \quad M, N = 1, \ldots, 248 \,,
\end{equation}
with appropriate fermionic fields \cite{Baguet:2016jph} in its supersymmetric completion, which will not be of importance to us here. Here $g_{\mu\nu}$ is the 3 dimensional metric and $\gM_{MN} \in \EE/\SO{16}$ the generalised metric containing all the fully internal bosonic fields. The gauge fields $A_\mu{}^M$ and $B_{\mu\,M}$ come from the bosonic fields with one external leg and transform in the $\mbf{248}$-dimensional adjoint representation of $\EE$.
	
The gauge structure of the $\EE$ ExFT is encoded via the generalised Lie derivative. However, unlike for higher-dimensional ExFTs, the parameters for the generalised Lie derivative consist not just of a generalised vector field, $\Lambda^M$, transforming in the $\mbf{248}$, but also of a ``covariantly constrained'' parameter $\Sigma_M$ transforming in the $\mbf{248}$. In terms of the parameters $\left(\Lambda^M,\, \Sigma_M\right)$, the generalised Lie derivative of a generalised vector field $V^M$ of weight $\lambda$ is given by
\begin{equation} \label{eq:GenLie}
	\gL_{(\Lambda,\Sigma)} V^M = \Lambda^N \partial_N V^M - 60\, \left(\mathbb{P}_{248}\right)^{M}{}_N{}^K{}_L V^N \partial_K \Lambda^L + \lambda\, V^M \partial_N \Lambda^N + f^{MN}{}_{K} \Sigma_N V^K \,.
\end{equation}
Here we have used the $\EE$ structure constants $f^{MN}{}_K$ and the projector onto the adjoint, $\mathbb{P}_{248}$, defined in \eqref{eq:ProjAdj}. The adjoint indices of $\EE$ are raised/lowered throughout with the Cartan-Killing metric, $\eta^{MN}$, normalised as in \eqref{eq:CKMetric}.
	
The derivatives $\partial_M$, the parameters $\Sigma_M$ appearing in the generalised Lie derivative \eqref{eq:GenLie} and the gauge fields $B_{\mu\,M}$ appearing in \eqref{eq:BosField}, are ``covariantly constrained'', meaning that
\begin{equation} \label{eq:CovConst}
	\left( \mathbb{P}_{1+248+3875} \right)_{MN}{}^{KL} C_K \otimes C'_L = 0 \,,
\end{equation}
for any $C_K$, $C'_K \in \left\{ \partial_M,\, \Sigma_M,\, B_{\mu\,M} \right\}$ and where the projectors are defined in \eqref{eq:ProjAdj} and \eqref{eq:Proj}. The derivatives are constrained in a similar way to \eqref{eq:CovConst} in higher-dimensional ExFTs, where the analogous constraint is known as the section condition. This implies that not all 248 coordinates appearing in $\partial_M$ are physical, but only a subset. There are two inequivalent maximal solutions to this section condition, where only 8 or 7 coordinates are kept, as we will review in further detail in \ref{s:SecCon}. The $\EE$ ExFT then reduces to 11-dimensional/IIB supergravity. By contrast, the constraints \eqref{eq:CovConst} on $\Sigma_M$ and $B_{\mu\,M}$ are new to $\EE$ ExFT and reflect the fact that to make the $\EE$ symmetry manifest, additional unphysical degrees of freedom transforming in the fundamental of $\EE$ are included in the field content \eqref{eq:BosField} which are removed by the additional symmetries associated to $\Sigma_M$ \cite{Hohm:2014fxa}.
	
The gauge structure parameterised by $\left(\Lambda^M,\,\Sigma_M\right)$ is more conveniently formulated in terms of the combined object \cite{Hohm:2017wtr}
\begin{equation}
	\Upsilon = \left( \Lambda,\,\Sigma \right) \,,
\end{equation}
where $\Lambda$ is a generalised vector field of weight 1 and $\Sigma$ is a covariantly constrained field of weight 0. The generalised Lie derivative can now be defined on such combined objects as follows
\begin{equation}
	\gL_{\Upsilon_1} \Upsilon_2 = \left( \gL_{\Upsilon_1} \Lambda_2{}^M ,\, \gL_{\Upsilon_1} \Sigma_{2\,M} + \Lambda_2{}^N \partial_M R_N(\Upsilon_1) \right) \,,
\end{equation}
with
\begin{equation}
	R^M(\Upsilon) = f^{MN}{}_K \partial_N \Lambda^K + \Sigma^M \,.
\end{equation}
	
This allows us to write an action for the bosonic fields of the $\EE$ ExFT as
\begin{equation}
	S = \int d^3 x\, d^{248} Y \sqrt{|g|} \left( \hat{R} + \frac1{240} g^{\mu\nu} D_\mu \gM_{MN} D_\nu \gM^{MN} + L_{\rm int}(\gM,g) + \frac{1}{\sqrt{|g|}} L_{CS} \right) \,,
\end{equation}
where $|g|$ denotes the determinant of the 3-dimensional metric $g_{\mu\nu}$. $\hat{R}$ its $\En{8}$-covariantised Ricci scalar, constructed as usual but replacing 3-dimensional partial derivatives $\partial_\mu$ by the 3-dimensional $\EE$-covariant derivatives $D_\mu$, defined as
\begin{equation}
	D_\mu = \partial_\mu - \gL_{(A_\mu,B_\mu)} \,,
\end{equation}
while
\begin{equation} \label{eq:Lint}
	\begin{split}
		L_{\rm int}(\gM,g) &= \frac1{240} \gM^{MN} \partial_M \gM^{KL} \partial_N \gM_{KL} - \frac12 \gM^{MN} \partial_M \gM^{KL} \partial_L \gM_{NK} \\
		& \quad - \frac1{7200} f^{NQ}{}_P f^{MS}{}_R \gM^{PK} \partial_M \gM_{QK} \gM^{RL} \partial_N \gM_{SL} \\
		& \quad + \frac12 \partial_M \ln |g| \partial_N \gM^{MN} + \frac14 \gM^{MN} \left( \partial_M \ln |g| \partial_n \ln|g| + \partial_M g^{\mu\nu} \partial_N g_{\mu\nu} \right) \,.
	\end{split}
\end{equation}
The Chern-Simons term is given by \cite{Hohm:2017wtr}
\begin{equation} \label{eq:LCS}
	L_{CS} = \epsilon^{\mu\nu\rho} \left( \langle \fA_\mu,\, \partial_\nu \fA_\rho \rangle - \frac13 \langle \fA_\mu,\, \gL_{\fA_\nu} \fA_\rho \rangle \right) \,,
\end{equation}
in terms of the 3-dimensional alternating tensor density $\epsilon^{\mu\nu\rho} = \pm1$ and the $\EE$ invariant inner product
\begin{equation} \label{eq:E8IP}
	\langle \fA_1,\, \fA_2 \rangle = \int d^{248}Y \left( A_1{}^M R_M(\fA_2) + A_2{}^M B_{1\,M} \right) \,,
\end{equation}
where $\fA = \left( A,\, B \right)$ denotes as before a generalised vector field of weight 1 and a covariantly constrained field of weight 0.
	
As usual, $L_{CS}$ is only gauge invariant up to a total derivative (in the external 3-dimensional spacetime). Equivalently, the Chern-Simons term can be written as an integral over four external spacetime dimensions but in a manifestly gauge-invariant manner in terms of the field strengths of $A_\mu{}^M$, $B_{\mu\,M}$, which can be computed from
\begin{equation} \label{eq:FieldStrengths}
	\left[ D_\mu,\, D_\nu \right] V^M = - \gL_{(\FF_{\mu\nu},\FG_{\mu\nu})} V^M \,,
\end{equation}
for any generalised vector field $V^M$. The Chern-Simons term is given in terms of $\FF_{\mu\nu}{}^M$ and $\FG_{\mu\nu\,M}$ by
\begin{equation} \label{eq:SCS}
	S_{CS} = \frac14 \int d^4x\, d^{248} Y \left( \FF^M \wedge \FG_M - \frac12 f_{MN}{}^K \FF^M \wedge \partial_K \FG^N \right) \,.
\end{equation}
The $\wedge$ in \eqref{eq:SCS} denotes the four-dimensional wedge product.

\subsection{Solutions of the section condition} \label{s:SecCon}
The dependence of the $\EE$ ExFT fields on the 248 coordinates is constrained by the ``section condition'' \eqref{eq:CovConst}
\begin{equation} \label{eq:SecCon}
	\left( \mathbb{P}_{1+248+3875} \right) _{MN}{}^{KL} \partial_K \otimes \partial_L = 0 \,,
\end{equation}
where the $\partial_M$ act on any single or product of fields of the $\EE$ ExFT. There are two inequivalent (up to $\EE$ transformations) maximal solutions of \eqref{eq:SecCon}, corresponding to 11-d and IIB supergravity.

A convenient way to solve the section condition to recover 11-d SUGRA comes by breaking $\EE \rightarrow \SL{9}$, such that
\begin{equation}
	\begin{split}
		\mbf{248} &\rightarrow \mbf{80}\+\mbf{84}\+\obf{84} \,, \\
		\mbf{3875} &\rightarrow \mbf{80} \oplus \mbf{1215} \oplus \mbf{240} \oplus \mbf{1050} \oplus \obf{240} \oplus \obf{1050} \,.
	\end{split}
\end{equation}
Decomposing further under $\SL{8}\times\mathbb{R}^+$, we obtain for the representations appearing in the $\mbf{248}$
\begin{equation} \label{eq:248SL8}
	\begin{split}
		\mbf{80}&\rightarrow\mbf{63}_0\+\mbf{1}_0\+\mbf{8}_3\+\obf{8}_{-3} \,, \\
		\mbf{84}&\rightarrow\mbf{56}_1\+\mbf{28}_{-2} \,, \\
		\obf{84}&\rightarrow \obf{56}_{-1}\+\obf{28}_{2} \,,
	\end{split}
\end{equation}
with the additional representations in the $\mbf{3875}$ decomposing as
\begin{equation} \label{eq:3875SL8}
	\begin{split}
		\mbf{240} &\rightarrow \mbf{36}_{-2} \oplus \mbf{168}_1 \oplus \mbf{8}_{-5} \oplus \mbf{28}_{-2} \,, \\
		\obf{240} &\rightarrow \obf{36}_{2} \oplus \obf{168}_{-1} \oplus \obf{8}_{5} \oplus \obf{28}_{2} \,, \\
		\mbf{1050} &\rightarrow \mbf{420}_{-2} \oplus \mbf{504}_{1} \oplus \mbf{56}_1 \oplus \mbf{70}_{4} \,, \\
		\obf{1050} &\rightarrow \obf{420}_{2} \oplus \obf{504}_{-1} \oplus \obf{56}_{-1} \oplus \obf{70}_{-4} \,, \\		
		\mbf{1215} &\rightarrow \mbf{216}_3 \oplus \mbf{63}_0 \oplus \mbf{720}_0 \oplus \obf{216}_{-3} \,.
	\end{split}
\end{equation}
We can now solve the section condition by restricting the coordinate dependence to lie solely in the $\obf{8}_{-3}$ of the $\mbf{80}$ of $\SL{9}$ \cite{Hohm:2014fxa}. Since the $\mbf{1} \oplus \mbf{248} \oplus \mbf{3875}$ do not contain any representations with charge $-6$ under $\mathbb{R}^+$, see \eqref{eq:248SL8}, \eqref{eq:3875SL8}, these 8 coordinates solve the section condition. In turn, $\EE$ ExFT reduces to 11-dimensional supergravity in a $3+8$ split.

An alternative solution \cite{Hohm:2014fxa} of the section condition comes by instead breaking $\SL{9} \rightarrow \SL{7} \times \SL{2} \times \mathbb{R}^+$, so that we have
\begin{equation} \label{eq:248SL7}
	\begin{split}
		\mbf{80} &\rightarrow \mbfb{48,1}_0 \oplus \mbfb{1,3}_0 \oplus \mbfb{1,1}_0 \oplus \mbfb{7,2}_3 \oplus \mbfb{\ov{7},2}_{-3} \,, \\
		\mbf{84} &\rightarrow \mbfb{7,1}_{-4} \oplus \mbfb{21,2}_{-1} \oplus \mbfb{35,1}_2 \,, \\
		\obf{84} &\rightarrow \mbfb{\ov{7},1}_{4} \oplus \mbfb{\ov{21},2}_{1} \oplus \mbfb{\ov{35},1}_2 \,,
	\end{split}
\end{equation}
for the representations in the $\mbf{248}$ and
\begin{equation} \label{eq:3875SL7}
	\begin{split}
		\mbf{240} &\rightarrow \mbfb{28,2}_{-1} \oplus \mbfb{112,1}_{2} \oplus \mbfb{7,3}_{-4} \oplus \mbfb{7,1}_{-4} \oplus \mbfb{21,2}_{-1} \oplus \mbfb{1,2}_{-7} \,, \\
		\obf{240} &\rightarrow \mbfb{\ov{28},2}_{1} \oplus \mbfb{\ov{112},1}_{-2} \oplus \mbfb{\ov{7},3}_{4} \oplus \mbfb{\ov{7},1}_{4} \oplus \mbfb{\ov{21},2}_{1} \oplus \mbfb{1,2}_{7} \,, \\
		\mbf{1050} &\rightarrow \mbfb{140,1}_{-4} \oplus \mbfb{224,2}_{-1} \oplus \mbfb{21,2}_{-1} \oplus \mbfb{210,1}_{2} \oplus \mbfb{35,3}_{2} \oplus \mbfb{35,1}_{2} \oplus \mbfb{\ov{35},2}_{5} \,, \\
		\obf{1050} &\rightarrow \mbfb{\ov{140},1}_{4} \oplus \mbfb{\ov{224},2}_{1} \oplus \mbfb{\ov{21},2}_{1} \oplus \mbfb{\ov{210},1}_{-2} \oplus \mbfb{\ov{35},3}_{-2} \oplus \mbfb{\ov{35},1}_{-2} \oplus \mbfb{35,2}_{-5} \,, \\		
		\mbf{1215} &\rightarrow \mbfb{\ov{140},2}_{-3} \oplus \mbfb{392,1}_{0} \oplus \mbfb{48,3}_{0} \oplus \mbfb{48,1}_{0} \oplus \mbfb{140,2}_{3} \oplus \mbfb{7,2}_{3} \oplus \mbfb{21,1}_{6} \\
		& \quad \oplus \mbfb{\ov{21},1}_{-6} \oplus \mbfb{\ov{7},2}_{-3} \oplus \mbfb{1,1}_{0} \,.
	\end{split}
\end{equation}
for the additional representations in the $\mbf{3875}$. Again, we can solve the section condition by allowing only coordinate dependence on the $\mbfb{\ov{7},1}_4$, since the $\mbf{1} \oplus \mbf{248} \oplus \mbf{3875}$ do not contain any representations with charge $+8$ under $\mathbb{R}^+$, as can be seen from \eqref{eq:248SL7} and \eqref{eq:3875SL7}. The $\EE$ ExFT then reduces to IIB supergravity in a $3+7$ split.

\section{Generalised Killing Vector fields} \label{s:GenKill}
We begin by establishing some facts about the action of generalised Killing vector fields, since these play a key role in consistent truncations. Just as in higher-dimensional ExFTs, we wish to define a generalised Killing vector as one which annihilates the generalised metric under the generalised Lie derivative. However, in $\EE$ ExFT we also need to specify a covariantly-constrained field $\Sigma$. Therefore, we say that $\left(V^M, \Sigma_M\right)$ is a generalised Killing vector field if
\begin{equation} \label{eq:GenKilVec}
	\gL_{(V, \Sigma)} \gM_{MN} = 0 \,.
\end{equation}
Using an explicit parameterisation of the generalised metric \cite{Hohm:2014fxa}, we can now determine what \eqref{eq:GenKilVec} implies for the components of $V$ and $\Sigma$. Let us label by $i, j = 1, \ldots, n$, the coordinates satisfying the section condition, with $n = 8$ or $n = 7$, depending on whether we are looking at a solution of the section condition corresponding to 11-d or type IIB supergravity. Consider now the rescaled generalised metric $\tilde{\gM}^{MN} = \gM^{MN} |g|^{1/3}$, where $|g|$ refers to the determinant of the external 3-dimensional metric. Then the following component of this rescaled generalised metric
\begin{equation}
	\tilde{\gM}^{ij} = h^{ij} \,,
\end{equation}
always encodes the inverse of the internal spacetime metric, $h_{ij}$. Moreover, the generalised Lie derivative on this component always reduces to the action of the ordinary Lie derivative
\begin{equation}
	\gL_{(V,\Sigma)} \left( \tilde{\gM}^{ij} \right) = L_{v} h^{ij} \,,
\end{equation}
where $v$ denotes the vector field component of $V^M$ and $L_{v}$ is the ordinary Lie derivative with respect to the vector field $v$. Therefore, for a generalised Killing vector field, the vector component corresponds to an ordinary Killing vector:
\begin{equation} \label{eq:GenKilVecC1}
	L_{v} h^{ij} = 0 \,.
\end{equation}
Moreover, by studying the remaining components of the generalised Killing vector action, we see that on the $(p+1)$-form potentials, $C_{(p+1)}$, we must have
\begin{equation} \label{eq:GenKilVecC2}
	L_{v} C_{(p+1)} = d\lambda_{(p)} \,,
\end{equation}
where $\lambda_{(p)}$ correspond to the $p$-form components of the generalised vector fields $V^M$. Therefore, the Killing vector fields $v$ are isometries of the background which also preserve the field strengths of the background, just as in higher dimensions.

Moreover, since $\Sigma$ is covariantly constrained as in \eqref{eq:CovConst}, the only non-vanishing part of $\Sigma$ is the 1-form. Using \eqref{eq:GenKilVecC1} and \eqref{eq:GenKilVecC2}, the generalised Killing vector condition \eqref{eq:GenKilVec} now imposes
\begin{equation} \label{eq:GenKilVecC3}
	\Sigma_{i} = \partial_{j} \tau^{j}{}_i + \partial_{i} \tau \,,
\end{equation}
where $\tau^i{}_j$ and $\tau$ parameterise the adjoint- and singlet-valued components of the generalised vector field $V$, respectively. This can, for example, be seen by looking at the following component when solving the section condition in a way that recovers 11-dimensional supergravity:
\begin{equation} \label{eq:SigmaCondition}
	\begin{split}
			\left(\gL _{(V,\Sigma)} \tilde{\gM}\right)^{i,j}{}_k &= \PR{63}^j{}_k{}^l{}_m h^{im} \left(\Sigma_{l} - \partial_{n} \tau^n{}_l - \partial_{l} \tau \right) + \frac38 \delta^j{}_k h^{il} \left(\Sigma_{l} - \partial_{m} \tau^{m}{}_l - \partial_{l} \tau \right).
		\end{split}
\end{equation}
Requiring \eqref{eq:SigmaCondition} to vanish, as for a generalised Killing vector field, immediately yields \eqref{eq:GenKilVecC3}, and a completely analogous consideration lead to \eqref{eq:GenKilVecC3} in IIB. Note that $\Sigma_i$, $\tau^i{}_j$ and $\tau$ only ever appear in the generalised Lie derivative through the combination $(\Sigma_i - \partial_j \tau^j{}_i - \partial_i \tau)$. Therefore, \eqref{eq:GenKilVecC3} implies that $\Sigma_i$, $\tau^i{}_j$ and $\tau$ necessarily drop out of the action of the generalised Lie derivative for generalised Killing vector fields. Finally, the remaining parts of the generalised vector field, which also do not have a clear geometric origin, drop out of the action of the generalised Lie derivative.

We now note that if we have a generalised Killing vector field $\left(V^t,\Sigma^t\right)$, whose vector part vanishes identically, i.e. $v = 0$, then \eqref{eq:GenKilVecC2} implies that the $p$-forms in $V^M$ must be closed. As a result, the action of $\left(V^t,\Sigma^t\right)$ under the generalised Lie derivative is trivial on any tensor. Therefore, we identically have
\begin{equation}
	\gL_{\left(V^t,\Sigma^t\right)} = 0 \,,
\end{equation}
when acting on any generalised tensor. We call such $\left(V^t,\Sigma^t\right)$ trivial generalised Killing vector fields. 

\section{Generalised Scherk-Schwarz reduction} \label{s:GSS}
We will now show how to construct consistent truncations to 3-dimensional gauged supergravity which preserve all supersymmetries. In higher-dimensional ExFTs, such consistent truncations arise from a generalised Scherk-Schwarz ansatz \cite{Berman:2012uy,Berman:2013cli,Hohm:2014qga,Lee:2014mla}, consisting of an $E_{d(d)}$-valued matrix $U_{\ov{M}}{}^M$, known as a twist matrix, and a scalar density $\rho \in \mathbb{R}^+$ subject to a differential condition. The twist matrix $U_{\ov{M}}{}^M$ and scalar field $\rho$ can be used to construct a set of generalised vector fields ${\cal U}_{\ov{M}}{}^M = \rho^{-1} U_{\ov{M}}{}^M$. In terms of these generalised vector fields, the differential condition is given by
\begin{equation} \label{eq:GenParallel}
	\gL_{{\cal U}_{\ov{M}}} {\cal U}_{\ov{N}}{}^M = X_{\ov{M}\ov{N}}{}^{\ov{P}} {\cal U}_{\ov{P}}{}^M \,,
\end{equation}
with $X_{\ov{M}\ov{N}}{}^{\ov{P}}$ constant. The condition \eqref{eq:GenParallel} implies that the manifold on which the truncation is performed is generalised Leibniz parallelisable \cite{Lee:2014mla}.
	
Similarly, consistent truncations to 3-dimensional maximal gauged SUGRA are captured in $\EE$ ExFT by a generalised Scherk-Schwarz ansatz parameterised by a twist matrix $U_{\ov{M}}{}^M \in \EE$ and a scalar field $\rho \in \mathbb{R}^+$. However, the generalised parallelisable condition \eqref{eq:GenParallel} is modified since the gauge structure of $\EE$ requires a covariantly constrained field in addition to a generalised vector field. Therefore, we need to also specify a covariantly constrained field in order to define a consistent truncation. In a similar way to consistent truncations in $\ON{d+1,d+1}$ double field theory \cite{Hohm:2017wtr}, such a covariantly constrained field can actually be constructed from the twist matrix $U_{\ov{M}}{}^M$ and $\rho$ as
\begin{equation} \label{eq:CovConstT}
	\Sigma_{\ov{M}M} = \frac{1}{60} \rho^{-1} f_{\ov{M}}{}^{\ov{P}\ov{Q}} U_{\ov{P}P} \partial_M U_{\ov{Q}}{}^P = \frac{1}{60} \rho^{-1} f_{\ov{M}}{}^{\ov{P}\ov{Q}} \mathrm{Tr} \left( U_{\ov{P}} \partial_M U_{\ov{Q}} \right) \,.
\end{equation}
The generalised Leibniz parallelisability condition can now be expressed in terms of
\begin{equation} \label{eq:DoubleU}
	\ff{U}_{\ov{M}} = ({\cal U}_{\ov{M}},\,\Sigma_{\ov{M}}) \,,
\end{equation}
with $\Sigma_{\ov{M}}$ given in \eqref{eq:CovConstT} and ${\cal U}_{\ov{M}}{}^M = \rho^{-1} U_{\ov{M}}{}^M$, as
\begin{equation} \label{eq:GenParallCondition}
	\gL_{\ff{U}_{\ov{M}}} {\cal U}_{\ov{N}}{}^M = X_{\ov{M}\ov{N}}{}^{\ov{P}} {\cal U}_{\ov{P}}{}^M \,.
\end{equation}
	
However, due to the form of $\Sigma_{\ov{M}}$ and the coefficient $\frac{1}{60}$ in \eqref{eq:CovConstT}, the generalised Leibniz parallelisability condition \eqref{eq:GenParallCondition} furthermore implies that the full $\ff{U}_{\ov{M}} = \left({\cal U}_{\ov{M}},\, \Sigma_{\ov{M}}\right)$ forms an algebra under the generalised Lie derivative:
\begin{equation} \label{eq:GenParallFull}
	\gL_{\ff{U}_{\ov{M}}} \ff{U}_{\ov{N}} = X_{\ov{M}\ov{N}}{}^{\ov{P}} \ff{U}_{\ov{P}} \,,
\end{equation}
with the additional condition in \eqref{eq:GenParallFull}, i.e.
\begin{equation}\label{eq:GenParallSigma}
	\gL_{({\cal U}_{\ov{M}}, \Sigma_{\ov{M}})} \Sigma_{\ov{M}\,M} + {\cal U}_{\ov{N}}{}^N \partial_M R_N({\cal U}_{\ov{M}}, \Sigma_{\ov{M}}) = X_{\ov{M}\ov{N}}{}^{\ov{P}} \Sigma_{\ov{P}\,M} \,,
\end{equation}
automatically satisfied.
	
Equation \eqref{eq:GenParallFull} plays an important role in constructing a consistent truncation, since it allows us to expand the constrained field $B_{\mu\,M}$ of $\En{8}$ ExFT in terms of $\Sigma_{\ov{M}M}$. Another important feature of the condition \eqref{eq:GenParallCondition} with \eqref{eq:CovConstT} is that the constants $X_{\ov{M}\ov{N}}{}^{\ov{P}}$ satisfy the linear constraint of maximal 3-dimensional gauged supergravities \cite{Nicolai:2000sc,Nicolai:2001sv}, i.e. that $X_{\ov{M}\ov{N}}{}^{\ov{P}}$ only lives in the representations
\begin{equation} \label{eq:LinConstraint}
	X_{\ov{M}\ov{N}}{}^{\ov{P}} \in \mathbf{1} \oplus \mathbf{248} \oplus \mathbf{3875} \,.
\end{equation}
In particular, the linear constraint \eqref{eq:LinConstraint} is automatically satisfied if we have a twist matrix $U_{\ov{M}}{}^M$ and scalar density $\rho$, due to the form of $\Sigma_{\ov{M}}$ and the precise coefficient $\frac{1}{60}$ in \eqref{eq:CovConstT}.

\subsection{Truncation ansatz}
We can now give the consistent truncation ansatz for the generalised Scherk-Schwarz reduction. For this, we simply expand the $\En{8}$ fields in terms of the generalised frame ${\cal U}_{\ov{M}}{}^M$, $\rho$ and $\Sigma_{\ov{M}}$ as follows
\begin{equation}
	\begin{split}
		g_{\mu\nu}(x,Y) &= \rho^{-2}(Y)\, G_{\mu\nu}(x) \,, \\
		{\cal M}_{MN}(x,Y) &= U^{\ov{M}}{}_M(Y)\, U^{\ov{N}}{}_N(Y) \, \gM_{\ov{M}\ov{N}} \,, \\
		A_\mu{}^M(x,Y) &= {\cal U}_{\ov{M}}{}^M(Y)\, {\cal A}_\mu{}^{\ov{M}}(x) \,, \\
		B_{\mu\,M}(x,Y) &= \Sigma_{\ov{M}\,M}(Y)\, {\cal A}_{\mu}{}^{\ov{M}}(x) \,.
	\end{split} \label{eq:TruncAnsatz}
\end{equation}
The condition \eqref{eq:GenParallFull} now ensures that the $Y$-dependence of all these objects factorises under the generalised Lie derivative. In particular, we have that
\begin{equation}
	\begin{split}
		D_{\mu} g_{\nu\rho}(x,Y) &= \rho^{-2}(Y) \left( \partial_\mu G_{\nu\rho}(x) - {\cal A}_\mu{}^{\ov{M}}(x)\, \theta_{\ov{M}}\, G_{\nu\rho}(x) \right) \equiv \rho^{-2}(Y)\, {\cal D}_{\mu} G_{\nu\rho}(x) \,, \\
		D_{\mu} \gM_{MN}(x,Y) &= U_{M}{}^{\ov{M}}(Y)\, U_N{}^{\ov{N}}(Y) \left( \partial_\mu \gM_{\ov{M}\ov{N}}(x) - 2\, {\cal A}_{\mu}{}^{\ov{P}}(x)\, X_{\ov{P}(\ov{M}}{}^{\ov{Q}} \gM_{\ov{N})\ov{Q}}(x) \right) \\
		&\equiv U_{M}{}^{\ov{M}}(Y)\, U_N{}^{\ov{N}}(Y) {\cal D}_{\mu} \gM_{\ov{M}\ov{N}}(x) \,, \\
		\gL_{\ff{A}_\mu} \ff{A}_{\nu}(x,Y) &= \ff{U}_{\ov{P}}(Y)\, X_{\ov{M}\ov{N}}{}^{\ov{P}}\, {\cal A}_\mu{}^{\ov{M}}(x)\, {\cal A}_\nu{}^{\ov{N}}(x) \equiv \ff{U}_{\ov{M}}(Y)\, \Leib{{\cal A}_{\mu}}{{\cal A}_{\nu}}^{\ov{M}}(x) \,.
	\end{split}
\end{equation}
This ensures that the above truncation ansatz \eqref{eq:TruncAnsatz} is consistent and we obtain a three-dimensional gauged supergravity with embedding tensor $X_{\ov{M}\ov{N}}{}^{\ov{P}}$ defined by \eqref{eq:GenParallCondition}.
	
\subsection{Largest possible compact gauge group} \label{s:CompGauge}
Equation \eqref{eq:GenParallCondition} links the gauging, encoded in the embedding tensor $X_{\ov{M}\ov{N}}{}^{\ov{P}}$, to the generalised Lie derivative acting on the internal space. As a result, the $\EE$ ExFT geometry imposes restrictions on which gauged supergravities can arise from consistent truncations. In particular, as we will now show, the largest compact subgroup of the gauging is $\SO{9}$. While we focus here on maximally supersymmetric truncations, the same argument also holds for less supersymmetric truncations, so that also for these the largest compact gauging is $\SO{9}$.

To begin, let us denote the generalised vector fields associated to compact generators by $\ff{U}_{\ov{I}} = \left( {\cal U}_{\ov{I}},\, \Sigma_{\ov{I}} \right)$, such that
\begin{equation}
	\gL_{\ff{U}_{\ov{I}}} {\cal U}_{\ov{M}}{}^M = X_{\ov{I}\ov{M}}{}^{\ov{N}}\, {\cal U}_{\ov{N}}{}^M \,,
\end{equation}
and
\begin{equation}
	X_{\ov{I}(\ov{M}}{}^{\ov{P}}\, \delta_{\ov{N})\ov{P}} = 0 \,.
\end{equation}
As a result, these ``compact'' generalised vector fields leave invariant the generalised metric $\gM_{MN} = U_M{}^{\ov{M}}\, U_N{}^{\ov{N}}\, \delta_{\ov{M}\ov{N}}$, since
\begin{equation}
	\begin{split}
		\gL_{\ff{U}_{\ov{I}}} \gM_{MN} &= 2\, U_{(M}{}^{\ov{M}}\, \delta_{\ov{M}\ov{N}}\, \gL_{\ff{U}_{\ov{I}}} U_{N)}{}^{\ov{N}} \\
		&= - 2\, U_{M}{}^{\ov{M}}\, U_{N}{}^{\ov{P}}\, X_{\ov{I}(\ov{P}}{}^{\ov{N}}\, \delta_{\ov{M})\ov{N}} \\
		&= 0 \,.
	\end{split}
\end{equation}
Thus, the $\ff{U}_{\ov{I}}$ are generalised Killing vector fields, as discussed in \ref{s:GenKill}.

We thus see that the compact gauging is generated by Killing vector fields of the background which preserve the field strengths as well. Moreover, the gauging is always realised by the vector fields, i.e. if we denote by $i = 1, \ldots, n$ the internal coordinates satisfying the section condition, with $n = 8$ or $7$ depending on whether we are looking at the 11-d or type II supergravity solutions, then we have
\begin{equation} \label{eq:VecGauging}
	\gL_{\ff{U}_{\ov{M}}} \ff{U}_{\ov{N}}{}^i = L_{v_{\ov{M}}} v_{\ov{N}}{}^i = X_{\ov{M}\ov{N}}{}^{\ov{P}}\, v_{\ov{P}}{}^i \,.
\end{equation}
In particular, this means that the vector fields $v_{\ov{M}}$ are valued in the adjoint of the gauge group. Moreover, the compact gauging is realised by the Lie bracket of Killing vectors of the backgrounds. However, there can be at most $\frac12 n(n+1)$ Killing vector fields on a $n$-dimensional manifold, which would then have to be a maximally symmetric space, in this case the $S^8$ or $S^7$. The trivial generalised Killing vector fields, which have vanishing vector components, have a trivial action under the generalised Lie derivative and, therefore, cannot contribute to the gauging. As a result, the largest possible compact gauging is $\SO{9}$, which would have to be a $S^8$ for 11-d supergravity, and $\SO{8}$, corresponding to a $S^7$ for IIB supergravity.

However, it should be noted that $\SO{9}$ cannot be gauged in three dimensions, because the embedding tensor representations \eqref{eq:LinConstraint} do not contain $\SO{9}$ singlets. On the other hand, $\SO{8}$ is realised in both IIA and IIB supergravity via a consistent truncation of $S^7$ leading to two different 3-dimensional $\SO{8} \ltimes T^{28}$ gauged supergravity, as we will explicitly construct in section \ref{s:S7}.

Since we have shown that the largest compact group that can be gauged is $\SO{9}$, this immediately rules out large numbers of ${\cal N}=16$ gauged SUGRAs. In particular, three-dimensional gauged supergravity admits large gaugings, such as all of $\EE$, different real forms of $\SO{8} \times \SO{8}$, and so on \cite{Nicolai:2001sv}. Indeed, none of the gauged supergravities analysed in \cite{Nicolai:2001sv} can arise from consistent truncations of 10-/11-dimensional supergravity.

Moreover, our argument for the largest compact gauge group just relies on the fact that the gauging is realised by the vector fields, i.e. \eqref{eq:VecGauging}. This is also true for less than maximally supersymmetric consistent truncations, which we discuss in \ref{s:LessSUSY}. Therefore, also for ${\cal N} < 16$ gauged SUGRAs, only those whose compact gauging is smaller than $\SO{9}$ can be uplifted by a consistent truncation to 10-/11-dimensional supergravity. This strong results rules out, for example, almost all the half-maximal gaugings with ${\cal N}=(8,0)$ AdS$_3$ vacua constructed in \cite{Deger:2019tem}.

\subsection{No-go theorem for truncations on products of spheres}\label{s:ProdTruncation}
We can use the fact that the vector fields generate the gauging, see \eqref{eq:VecGauging}, and therefore are valued in the adjoint representation of the gauge group, to rule out further consistent truncations. For example, in four dimensions, an interesting class of gauged supergravities are those with ``dyonic gaugings'' that arise from consistent truncations of type II supergravity on products of two spheres $S^p \times S^{6-p}$, as well as by replacing one or both of the spheres by hyperboloids \cite{Inverso:2016eet}. This is not possible in higher dimensions, but requires the electric-magnetic duality inherent to four dimensions. In the ExFT language, it is related to the fact that in the decomposition of $\En{7} \rightarrow \SL{8}$, the $\mathbf{56}$ coordinates decompose into the $\mbf{28} \oplus \obf{28}$. Under $\SL{8} \rightarrow \SL{p} \times \SL{8-p}$, the $\mbf{28}$ and $\obf{28}$ naturally contain the $\SO{p}$ and $\SO{8-p}$ adjoints. Moreover, and crucially, we can solve the section condition with $p-1$ coordinates that are part of the adjoint of $\SO{p}$ coming from the $\mbf{28}$ and $7-p$ coordinates that are part of the adjoint of $\SO{8-p}$ coming from the $\obf{28}$. Therefore, it is possible to have vector fields (which are dual to the coordinate representations and therefore also satisfy the section condition) transforming in the adjoint of $\SO{p} \times \SO{8-p}$. The same logic holds for non-compact gaugings. We refer the reader to \cite{Inverso:2016eet} for more details on the construction of the gauging.

A natural question is whether we can similarly define consistent truncations to 3-dimensional gauged supergravities on the 7-/8-dimensional product of several spheres, $S^p \times S^q \times \ldots$, leading to the gauging of the isometry groups $G = \SO{p+1} \times \SO{q+1} \times \ldots$, or hyperboloids with the respective non-compact gaugings. Here we are strictly interested for each sphere factor the full isometry groups is gauged, and thus exclude the case where, for example, we have a trivial $S^1$ reduction that does not lead to an additional $\SO{2}$ gauging. We will show here that these types of product truncations to three-dimensions are not consistent.

Let us focus on the spherical case. Since the gauging is compact, it is necessarily a subgroup of $\SO{16}$, the maximal compact subgroup of $\EE$. To have a consistent truncation on a product of spheres, such that all of the isometries are gauged, we need all of the vector fields to transform in the adjoint of the isometry groups, i.e. $G = \SO{p} \times \SO{q} \times \ldots \subset \SO{16}$. However, just like the coordinates on the products of the spheres, the vectors must satisfy the section condition of the $\EE$ theory. In particular, this means that if we consider a compactification on $S^{p} \times S^{7-p}$, a subset of the adjoints of $\SO{p+1}$ and $\SO{8-p}$, must satisfy the section condition. This subset corresponds precisely to $\mathfrak{so}(p+1) \ominus \mathfrak{so}(p)$ and $\mathfrak{so}(8-p) \ominus \mathfrak{so}(7-p)$, which is in one-to-one correspondence with the coordinates on $S^p \times S^{7-p}$. This extends in the obvious way to products of more than two spheres.

Let us now study the section condition of the $\EE$ ExFT with respect to $\SO{16} \subset \EE$. Recall that the section conditions \eqref{eq:SecCon} requires that the product of coordinates in the $\mbf{1} \oplus \mbf{248} \oplus \mbf{3875}$ must vanish. Under $\SO{16} \subset \EE$, the $\mbf{3875}$, in particular, decomposes as
\begin{equation}
	\mbf{3875} \rightarrow \mbf{135} \oplus \mbf{1920'} \oplus \mbf{1820} \,.
\end{equation}
Importantly, the $\mbf{1820}$ corresponds to the totally antisymmetric product of four fundamentals of $\SO{16}$. Therefore, for any product of subgroups of $\SO{16}$, the $\mbf{1820}$ will always contain the tensor product of the adjoint of two of these subgroups. As a result, we see that the section condition is never satisfied by vectors which transform as the adjoint of the gauging $G = \SO{p} \times \SO{q} \times \ldots \subset \SO{16}$, and therefore such product truncations are not possible. The only way to avoid this no-go theorem is by placing the vectors in the adjoint of the gauge group $G$ but also transforming in a non-trivial representation of the commutant of the gauge group $G$ inside $\SO{16}$. There are only a handful of ways of doing this, by considering gaugings within $\SU{8} \times \U{1} \subset \SO{16}$, $\USp{8} \times \SU{2} \subset \SO{16}$ or $\USp{4} \times \USp{4} \subset \SO{16}$. However, it is straightforward to study these cases individually and to see that the section condition always forbids having vector fields transforming in adjoint representations of $G$ such that $G$ corresponds to the full isometry group of a 7- or 8-dimensional product of spheres. Therefore, we conclude that there are no consistent truncations on a product of spheres from 10/11 dimensions to 3 dimensions in which the full isometry group is gauged.

Our no-go theorem can likely be extended to non-compact gaugings corresponding to a product of hyperboloids, possibly with some spheres, by considering appropriate gaugings inside $\SO{8,8} \subset \SO{16}$.
	
\section{$S^7$ and $H^{p,q}$ truncations of IIA/IIB} \label{s:S7}
Using the formalism described in section \ref{s:GSS}, we can now construct the consistent truncation of IIA and IIB supergravity on $S^7$ and hyperboloids $H^{p,q}$. In the case of $S^7$, the truncation leads to two different 3-dimensional $\SO{8} \ltimes T^{28}$ gauged supergravities, with $T^{28}$ the group of 28 translations. In this case, the $\SO{8}$ is embedded differently in $\EE$ for the IIA/IIB case \cite{Fischbacher:2003yw}. Otherwise, the truncations lead to the same 3-dimensional theories with gauging $\mathrm{CSO}(p,q,8-p-q) \ltimes T_{p,q,8-p-q}$, where $T_{p,q,8-p-q}$ is a group of translations of dimension $\frac12 (15-p-q)(p+q)$.

These consistent truncations can be constructed in terms of two different $\SL{8}$ subgroups of $\EE$, which we will denote as $\SLA$ and $\SLB$. The $\SLA$ is embedded in $\EE$ as follows:
\begin{equation}
	\EE \rightarrow \En{7} \times \SL{2} \rightarrow \SLA \times \SL{2} \,,
\end{equation}
with the $\mathbf{248}$ decomposing under $\SLA \times \SL{2}$ as
\begin{equation} \label{eq:SL8IIA}
	\mathbf{248} \rightarrow \mbfb{28,2} \oplus \mbfb{\ov{28},2} \oplus \mbfb{63,1} \oplus \mbfb{1,3} \oplus \mbfb{70,1} \,.
\end{equation}
On the other hand, the $\SLB$ is embedded as
\begin{equation}
	\EE \rightarrow \SL{9} \rightarrow \SLB \times \mathbb{R}^+ \,,
\end{equation}
with the $\mathbf{248}$ decomposing as
\begin{equation} \label{eq:SL8IIB}
	\mbf{248} \rightarrow \obf{8}_{-3} \oplus \mbf{28}_{-2} \oplus \obf{56}_{-1} \oplus \mbf{63}_0 \oplus \mbf{1}_0 \oplus \mbf{56}_1 \oplus \obf{28}_2 \oplus \mbf{8}_3 \,.
\end{equation}

For both the $\SLA$ and $\SLB$, we can solve the section condition as follows. Let the coordinates lie in just one of the $\mbf{28}$'s and let us denote these by $Y^{IJ}$ with $I, J = 0, \ldots, 7$. Note that for $\SLA$, this requires breaking the $\SL{2}$ commutant to select one of the doublets of $\mbf{28}$. Then, the section condition reduces on the chosen $\mbf{28}$ to
\begin{equation}
	\partial_{[IJ} \otimes \partial_{KL]} = 0 \,,
\end{equation}
which can be solved by identifying the 7 IIA/IIB physical coordinates as $Y^{i0}$, with $i = 1, \ldots, 7$.

In doing this for $\SLA$, we are picking the 7 coordinates as the standard ``11-dimensional'' solution of the section condition inside the $\mbf{56}$ of the $\En{7}$ subgroup. By restricting to just these 7 coordinates in the $\EE$ theory, it is clear that we have a IIA solution of the section condition and that fields can depend on one further coordinate while still obeying the section condition. This additional coordinate is in the $\trep{1,3}$ representation, specifically if one breaks $\SL{2}\rightarrow\mathbb{R}^+$, such that the $\mbf{2} = (+, -)$, the eighth coordinate can then be taken to be either $Y^{++}$ or $Y^{--}$ depending on whether the other $7$ coordinates live in the $\mbf{56}_+$, i.e. $Y^{{\cal A}\,+}$, or $\mbf{56}_-$, i.e. $Y^{{\cal A}\,-}$, respectively, where ${\cal A} = 1, \ldots, 56$ of $\En{7}$.

On the other hand, for $\SLB$ we are considering the same $\SL{8}$ decomposition as when we discussed the 11-dimensional solution to the section condition in \ref{s:SecCon}. However, unlike for the 11-dimensional solution of the section condition, we are now solving the section condition with antisymmetric coordinates in the $\mbf{28}_{-2}$ coming from the 3-vector coordinates in the $\mbf{84}$ of $\SL{9}$. This allows us to keep at most 7 coordinates whilst obeying the section condition. By contrast, the 8 coordinates corresponding to the 11-dimensional solution to the section condition sit inside the adjoint of $\SL{9}$.

Now that we have chosen our 7 IIA/IIB coordinates as sitting inside a $\mbf{28}$ of $\SLA$ or $\SLB$, we can construct an ansatz for the twist matrix $U_{\ov{M}}{}^M$ and $\rho$. We do this by parameterising the $\EE$ twist matrix in terms of a $\SL{8}$ matrix $V_I{}^{\ov{I}}$, as well as a scalar function.

\subsection{$\SLA$ twist equations}
For $\SLA$, we follow the embedding $\SLA \subset \En{7} \times \SL{2} \subset \EE$, and parameterise the $\EE$ twist matrix in terms of the $\SL{8}$ matrix $V_I{}^{\ov{I}}$ and the diagonal $\SL{2}$ matrix
\begin{equation} \label{eq:SL2M}
	v_{\ov{i}}{}^i=
	\begin{pmatrix}
		\sigma & 0 \\
		0 & \sigma^{-1}
	\end{pmatrix} \,,
\end{equation}
with $i = 1, 2$ and $\ov{i} = 1, 2$. Moreover, we take the $\EE$ scalar density to be $\rho = \sigma^2$. This will be required for the trombone to vanish. To analyse the generalised parallelisability condition \eqref{eq:GenParallCondition}, let us decompose the embedding tensor representations $\mathbf{1} \oplus \mathbf{248} \oplus \mathbf{3875}$. Firstly, we find that the $\mathbf{1}$ automatically vanishes. Decomposing the $\mathbf{3875}$ and $\mbf{248}$ under $\SLA \times \mathbb{R}^+ \subset \En{7} \times \SL{2}$, we find that only the components
\begin{equation}
	\begin{split}
		\mbf{36}_+ \oplus \mbf{420}_+ &\subset \mbfb{912,2} \subset \mbf{3875} \,, \\
		\mbf{28}_+ &\subset \mbfb{56,2} \subset \mbf{248} \,,
	\end{split}
\end{equation}
are non-vanishing. Thus, the generalised parallelisability condition \eqref{eq:GenParallCondition} reduces to
\begin{equation} \label{eq:TwistEqns}
	\begin{split}
		\sigma^{-1} \partial_{IJ} \left( V^{-1} \right)_{(\ov{I}}{}^I \left( V^{-1} \right)_{\ov{J})}{}^J &= -\frac72 \theta_{\ov{I}\ov{J}} \,, \\
		\sigma^{-1} \left( V^{-1} \right)_{\ov{I}\ov{J}\ov{K}}{}^{IJK} \partial_{IJ} V_K{}^{\ov{L}} -\frac16 \sigma^{-1} \partial_{IJ}\left( V^{-1} \right)_{[\ov{I}\ov{J}}{}^{IJ} \delta_{\ov{K}]}{}^{\ov{L}} &= \theta_{\ov{I}\ov{J}\ov{K}}{}^{\ov{L}} \,, \\
		\sigma^{-1} \partial_{IJ} \left(V^{-1}\right)_{\ov{I}\ov{J}}{}^{IJ} - 2 \sigma^{-1} \left(V^{-1}\right)_{\ov{I}\ov{J}}{}^{IJ} \partial_{IJ} \ln\sigma &= \vartheta_{\ov{I}\ov{J}} \,.
	\end{split}
\end{equation}
Here $\theta_{\ov{I}\ov{J}}$ corresponds to the $\mbf{36}_+$, $\theta_{\ov{I}\ov{J}\ov{K}}{}^{\ov{L}}$ the $\mbf{420}_+$ and $\vartheta_{\ov{I}\ov{J}}$ the $\mathbf{28}_+$.

We note that \eqref{eq:TwistEqns} are precisely the $\SL{n}$ twist equations discussed in \cite{Hohm:2014qga}, with $n = 8$. In particular, we can use the form of $V_I{}^{\ov{I}}$ given in \cite{Hohm:2014qga} to constructing consistent truncations on $S^7$ and $H^{p,q}$ to three-dimensional gauged supergravities with gaugings $\SO{8} \ltimes T^{28}$ and $\mathrm{CSO}(p,q,8-p-q) \ltimes T_{p,q,8-p-q}$.

Finally, we can more generally construct a consistent truncation using the $\SL{2}$ ansatz \eqref{eq:SL2M} and any $\En{7}$-valued twist matrix (not necessarily $\SLA$-valued) satisfying the $\En{7}$ twist equations. This would correspond to dimensionally reducing a consistent truncation to 4-dimensional supergravity on $S^1$ to 3 dimensions. In particular, we can thus embed any of the 4-dimensional dyonic gaugings of \cite{Inverso:2016eet} into $\EE$.

\subsection{$\SLB$}
For $\SLB$, the $\EE$ twist matrix is similarly constructed from an $\SL{8}$ matrix $V_{\ov{I}}{}^I$ and scalar function $\varphi$, via the embedding $\EE \rightarrow \SL{9} \rightarrow \SLB \times \mathbb{R}^+$. Accordingly, we parameterise each power of $\mathbb{R}^+$ with $\varphi$, and construct each $\SL{8}$ representation from $V_{\ov{I}}{}^I$. This immediately causes the $\mbf{1}$ representation in the embedding tensor to vanish. 

The function $\varphi$ can be fixed by requiring the $\EE$ trombone to be proportional to $\vartheta_{\ov{IJ}}$ of the $\SL{8}$ twist equations described in \cite{Hohm:2014qga}. More specifically, we can start from the ansatz $\rho= \sigma^\lambda,\:\:\varphi=\sigma ^{\omega}$ where we take $\sigma$ to be the $\mathbb{R}^+$ function in the $\SL{8}$ ansatz of \cite{Hohm:2014qga} and $\lambda,\,\omega$ are arbitrary powers that we can solve for. Then, the only non-vanishing component of the $\mbf{248}$ representation is the $\mbf{28}$ of $\SL{8}$, given by
\begin{equation}
	\begin{split}
		\frac{1}{2}\Theta_{\ov{IJ
		}}&=\sigma^{-\lambda-6\omega}(\partial _{IJ}(V^{-1})_{\ov{IJ}}{}^{IJ}-(6\omega + 2\lambda)(V^{-1})_{\ov{IJ}}{}^{IJ}\partial _{IJ}\ln\sigma) \,.
	\end{split}
\end{equation}
In order to have $\Theta_{\ov{IJ}}\propto\vartheta_{\ov{IJ}}$, the trombone of the $\SL{8}$ twist equations \eqref{eq:TwistEqns}, we must take
\begin{equation}
	\lambda =2\,,\qquad \omega = -\frac1{6} \,.
\end{equation}
Substituting this ansatz into \eqref{eq:GenParallCondition}, we find that the embedding tensor in the $\mbf{3875}$ contains only the representations $\mbf{36} \subset \obf{240}$ and $\mbf{420} \subset \obf{1050}$ as $\SLB \subset \SL{9}$. Moreover, these equations reduce to precisely the $\SL{8}$ twist equations \eqref{eq:TwistEqns} of \cite{Hohm:2014qga}. As a result, we immediately have the consistent truncations of IIB supergravity on $S^7$ and $H^{p,q}$ to three dimensions with gaugings $\SO{8} \ltimes T^{28}$ and $\mathrm{CSO}(p,q,8-p-q) \ltimes T_{p,q,8-p-q}$.

\subsection{Relation of IIA/IIB truncations via an outer automorphism}\label{s:Outer}
Having explicitly constructed the truncations of IIA/IIB on $S^7$ in terms of two different $\SL{8}$ subgroups of $\EE$, we will now highlight a different aspect of the relationship between them. Firstly, the NS-NS sector of ten-dimensional supergravity admits a consistent truncation on $S^7$ to 3-dimensional ${\cal N}=8$ gauged supergravity \cite{Hohm:2017wtr}. This is a half-maximal subtruncation of the two truncations we constructed above. Let us now investigate how the IIA/IIB truncations differ from the perspective of this half-maximal subtruncation.

Therefore, let us consider the subgroup $\SO{8,8} \subset \EE$ controlling the half-maximal theory. Then we have the following decompositions
\begin{equation}
	\begin{split}
		\mbf{248} &\rightarrow \mbf{120} \oplus \mbf{128'} \,, \\
		\mbf{3875} &\rightarrow \mbf{135} \oplus \mbf{1820} \oplus \mbf{1920'} \,.
	\end{split}
\end{equation}
The half-maximal gauged supergravities only have the embedding tensor representations $\mbf{1} \oplus \mbf{120} \oplus \mbf{135} \oplus \mbf{1820}$ out of the representations lying in the $\mbf{1} \oplus \mbf{248} \oplus \mbf{3875}$ allowed by the maximal theory. In particular, the $S^7$ truncations lead to half-maximal embedding tensors valued in the $\mbf{135}$ and gives rise to a $\SO{8} \ltimes T^{28}$ gauging in the half-maximal theory.

Let us first understand the difference between the 3-dimensional theories obtained from IIA/IIB.  Generically, one would expect that by considering the same truncation in the maximal theory, we would find new couplings between vector fields and generators which are in the $\mbf{128'} \subset \mbf{248}$, and hence have a larger gauging in the maximal theory, and one that distinguishes between IIA/IIB. However, this is not the case here. Let us write the $\mbf{248}$ vector fields of the ${\cal N}=16$ theory as $A_\mu{}^M = \left(A_\mu{}^{AB},\, A_\mu{}^{\dot{I}}\right)$, with $A, B = 1, \ldots, 16$ labelling the fundamental and $\dot{I} = 1, \ldots, 128$ the $\mbf{128'}$ of $\SO{8,8}$, and similarly for the generators $t^M = \left( t^{AB},\, t^{\dot{I}} \right)$. The only possible coupling in the maximal theory to an embedding tensor in the $\mbf{135}$ of $\SO{8,8}$ is 
\begin{equation}
	A_\mu{}^M\, X_{MN}\, t^N = A_\mu{}^{AB}\, t^{C}{}_B\, \theta_{AC} \,,
\end{equation}
where we have lowered the $\SO{8,8}$ fundamental indices with the $\SO{8,8}$-invariant metric $\eta_{AB}$. We see explicitly that the coupling in the maximal theory does not involve either $A_\mu{}^{\dot I}$ or $t^{\dot{I}}$ and thus the ${\cal N}=16$ and ${\cal N}=8$ gauge groups coincide. In both cases, it is just $\SO{8} \ltimes T^{28}$. The fact that the gauge group does not change between the IIA/IIB truncations is different from what occurs in truncations of IIA/IIB on $S^3$ \cite{Malek:2015hma} and $S^3 \times H^{1,2}$ \cite{Malek:2017cle}.

If the gauge group is identical, then how do we see the difference between the 3-dimensional theories? This comes by looking at the decomposition of $\SO{8,8} \rightarrow \SL{8}$. As we discussed above, there are two different $\SL{8}$'s in $\EE$, leading to the IIA/IIB truncations. Since there is a unique $\SL{8} \times \mathbb{R}^+$ in $\SO{8,8}$, we have the common decomposition
\begin{equation}
	\mbf{120} \rightarrow \mbf{28}_2 \oplus \mbf{63}_0 \oplus \mbf{1}_0 \oplus \obf{28}_{-2} \,.
\end{equation}
However, the two different $\SL{8}$'s inside $\EE$ now manifest themselves by two different decompositions of the $\mbf{128'}$. With respect to $\SLA$, we have
\begin{equation} \label{eq:128A}
	\mbf{128'} \rightarrow \mbf{1}_4 \oplus \obf{28}_2 \oplus \mbf{70}_0 \oplus \mbf{28}_{-2} \oplus \mbf{1}_{-4} \,,
\end{equation}
while with respect to $\SLB$, we have
\begin{equation} \label{eq:128B}
	\mbf{128'} \rightarrow \mbf{8}_{-3} \oplus \mbf{56}_{-1} \oplus \obf{56}_1 \oplus \obf{8}_3 \,.
\end{equation}
This implies that the $\mbf{128'}$ has different decompositions under the $\SO{8}$ factor of the gauge group. Therefore, while the fields in the half-maximal theory couple identically to the generators of $\SO{8} \ltimes T^{28}$, the fields of the ${\cal N}=16$ theory, which also transform in the $\mbf{128'}$, will have different couplings to $\SO{8} \ltimes T^{28}$, depending on whether we are looking at the IIA or IIB truncation. Another way to see the difference between the ${\cal N}=16$ theories is to see the $\SO{8}$ gauge groups as different diagonal embeddings of $\SO{8} \subset \SO{8} \times \SO{8} \subset \SO{16}$ \cite{Fischbacher:2003yw}.

With respect to the $\SO{8,8}$ group, the different decompositions of the $\mbf{128'}$ under the $\SL{8} \subset \SO{8,8}$ in \eqref{eq:128A}, \eqref{eq:128B} can equivalently be viewed as arising from an outer automorphism of $\SO{8,8}$ which exchanges the $\mbf{128}$ with the $\mbf{128'}$. Therefore, the IIA/IIB truncations on $S^7$ are related by the outer automorphism of $\SO{8,8}$, analogous to similar consistent truncations of IIA/IIB on the same background in four and seven dimensions \cite{Malek:2015hma,Malek:2017cle}. Therefore, we can relate the IIA/IIB twist matrices by the outer automorphism of $\SO{8,8}$, which exchanges the $\mbf{128}$ and $\mbf{128'}$ representations. One way to do this would be to decompose the IIA/IIB twist matrices with respect to $\EE \rightarrow \SO{8,8}$ and identify the $\mbf{120}$ and $\mbf{128'}$ parts. Instead, since we have the construction of the twist matrices with regards to the $\SLA$ and $\SLB$ groups, let us show how the outer automorphism acts withing these $\SL{8}$'s and exchanges the appropriate representations.

Firstly, let us look at the $\EE$ coordinates, which transform in the $\mbf{120} \oplus \mbf{128'}$ of $\SO{8,8}$. The coordinates in the $\mbf{120}$ are constrained by the section condition to satisfy \cite{Hohm:2017wtr}
\begin{equation} \label{eq:SO88SecCon}
	\begin{split}
		\partial_{[AB} \otimes \partial_{CD]} = \partial_{AC} \otimes \partial_{B}{}^C = 0 \,.
	\end{split}
\end{equation}
We can solve \eqref{eq:SecCon} with 7 coordinates sitting in either the $\mbf{28}$, $\obf{28}$ or $\mbf{63}$ coordinates of $\SL{8} \subset \SO{8,8}$. If we write $I, J = 0, \ldots, 7$ for the $\SL{8}$ fundamental and $i = 1, \ldots, 7$, then these solutions correspond to taking the coordinates to be either one of the following choices
\begin{equation}
	\left\{ Y^{0i} \right\},\, \left\{ Y_{0i} \right\},\, \left\{ Y^0{}_i \right\},\, \left\{ Y_0{}^i \right\} \,.
\end{equation}
All of these choices are equivalent up to $\ON{8,8}$ transformations, but not $\SO{8,8}$. For example, if we let the coordinates be in the $\mbf{28}_2$, corresponding to $Y^{0i}$, (or equivalently $\obf{28}_{-2}$ and thus $Y_{0i}$) and use the decomposition of $\mbf{128'}$ according to $\SLA$ \eqref{eq:128A}, then we find that the section condition allows us include one of the singlets in $\mbf{128'}$ as an extra coordinate. On the other hand, if we had chosen the coordinates to be in the $\mbf{63}_0$, corresponding to $Y^{0}{}_i$ or $Y_0{}^i$, and taken the IIA decomposition \eqref{eq:128A}, we would have found that the section condition does not allow us to keep dependene on any further coordinate. Therefore, in the $\SLA$ decomposition \eqref{eq:128A}, the coordinates in the $\mbf{28}_2$ (or $\obf{28}_{-2}$) correspond to IIA coordinates, where the section condition allows us to keep dependence on one extra coordinate, while the coordinates in the $\mbf{63}_0$ correspond to IIB coordinates, with no extra coordinate allowed by the section condition. On the other hand, if we had chosen the $\SLB$ decomposition, the story would have been reversed: the coordinates in the $\mbf{28}_2$ would have been the IIB ones and those in the $\mbf{63}_0$ would have been IIA.

We can now describe the action of the outer automorphism of $\SO{8,8}$ in terms of the $\SLA$ and $\SLB$ subgroups. In order to do this, we break to the common subgroup $\SLA \supset \SL{7} \subset \SLB$. As discussed above, the outer automorphism should take the $\mbf{28}$ of $\SLA$ into the $\mbf{63}$ of $\SLB$ and vice versa. Moreover, it should swap the decompositions \eqref{eq:128A} and \eqref{eq:128B}. This is realised by the following transformation
\begin{equation} \label{eq:OuterAut}
	V_A \rightarrow V^A = \eta^{AB}\, V_B \,,
\end{equation}
and similarly on any other $\SO{8,8}$ tensor, with $\eta_{AB}$ the $\SO{8,8}$ invariant written in terms of the $\SL{7}$ subgroup as
\begin{equation}
	\eta_i{}^j = \delta_i{}^j \,, \qquad \eta^i{}_j = \delta^i{}_j \,, \qquad \eta_{00} = 1 \,, \qquad \eta^{00} = - 1 \,,
\end{equation}
with all the other elements vanishing. The effect of the transformation \eqref{eq:OuterAut} is to swap the $\mbf{7} \leftrightarrow \obf{7}$ of $\SL{7}$ everywhere. It is straightforward to check that this indeed swaps the representations in \eqref{eq:128A} and \eqref{eq:128B}. Finally, the transformation \eqref{eq:OuterAut} maps the IIA/IIB twist matrices into each other.
	
\section{Consistent truncations preserving less supersymmetry} \label{s:LessSUSY}
We will now show how to construct consistent truncations of type II and 11-dimensional supergravity to a 3-dimensional ${\cal N} \leq 16$ supergravity. The setup is similar as in higher dimensions \cite{Malek:2016bpu,Malek:2017njj,Cassani:2019vcl,Josse:2021put}, i.e. rather than having a globally well-defined frame ${\cal U}_{\ov{M}}{}^M \in \EE$, we have just a set of globally well-defined generalised vector fields stabilised by some $G \subset \EE$. This set of well-defined generalised vector fields therefore define a generalised $G$-structure in the $\EE$ ExFT. The number of spinors, transforming in the $\mbf{16}$ of $\SO{16} \subset \EE$, which are stabilised by $G$ defines the number of supersymmetries preserved by the truncation\footnote{Throughout this paper we are assuming that our manifold is Spin, so that there is no obstruction to lifting to the double cover $\SO{16}$ of $\SO{16}/\mathbb{Z}_2$, which is the true subgroup of $\EE$.}.

Since the generalised vector fields transform in the adjoint of $\EE$, the set of well-defined generalised vector fields defining the $G$-structure will transform in the adjoint representation of the commutant of $G \in \EE$, which we denote as $\CG$. Let us here label the adjoint representation of $\CG$ by $A = 1, \ldots, \textrm{dim}\, \CG$, and the set of well-defined generalised vector fields by
\begin{equation}
	\left\{ \cJ_{A}{}^M \right\} \,.
\end{equation}
Since the $\cJ_A{}^M$ are the generators of $\CG$, they satisfy the following algebraic relations
\begin{equation} \label{eq:AlgCond}
	\begin{split}
		\frac1{60} \tr\left(\cJ_A\, \cJ_B\right) &= \rho^{-2}\, \kappa_{AB} \,, \\
		\left[ \cJ_A,\, \cJ_B \right] &= \rho^{-1}\, f_{AB}{}^C\, \cJ_C \,,
	\end{split}
\end{equation}
where the trace is taken in $\EE$ and $\left[\,\,,\, \right]$ denotes the commutator. $\kappa_{AB}$ and $f_{AB}{}^C$ are the Cartan-Killing form and structure contants of $\CG$, respectively, and $\rho$ is an $\EE$ scalar density of weight $-1$, just as in the maximally supersymmetric case. With explicit $\EE$ indices, we can write \eqref{eq:AlgCond}
\begin{equation}
	\begin{split}
		\cJ_A{}^M\, \cJ_B{}^N\, \eta_{MN} & = \rho^{-2}\, \kappa_{AB} \,, \\
		f_{MN}{}^P\, \cJ_A{}^M\, \cJ_B{}^N &= \rho^{-1}\, f_{AB}{}^C\, \cJ_C{}^P \,,
	\end{split}
\end{equation}
where $\eta_{MN}$ and $f_{MN}{}^P$ are the Cartan-Killing form and structure constants of $\EE$. Note the factor of $\frac1{60}$ due to the conventions \eqref{eq:CKMetric}. The conditions \eqref{eq:AlgCond} are the less supersymmetric analogues of the condition that ${\cal U}_{\ov{M}}{}^M \in \EE$. Note that for consistent truncations that preserve some supersymmetry, $\CG$ is semi-simple, so that $\kappa_{AB}$ is invertible. Therefore, we will use $\kappa^{AB}$ and $\kappa_{AB}$ to raise and lower the adjoint $\CG$ indices $A, B$ freely.

The differential condition that ensures consistency of the truncation is now modified from the maximally supersymmetric case  \eqref{eq:GenParallCondition} to
\begin{equation} \label{eq:GenParallConditionLS}
	\gL_{\left(\cJ_{A},\Sigma_{A}\right)} \cJ_{B}{}^M = X_{AB}{}^C\, \cJ_C{}^M \,,
\end{equation}
where now $\Sigma_{A\,M}$ is defined in terms of $\cJ_A{}^M$ and $\rho$ as
\begin{equation} \label{eq:LessSUSYCF}
	\Sigma_{A\,M} = \frac1{60}\, \rho\, f_A{}^{BC} \tr\left( \cJ_B\, \partial_M \cJ_C \right) \,.
\end{equation}
For a consistent truncation, we must have that $X_{AB}{}^C$ is constant. Note that we can interpret \eqref{eq:GenParallConditionLS} as the condition that the intrinsic torsion of the $G$-structure contains only singlets under $G$, so that it is the $\EE$ analogue of this condition in higher-dimensional ExFTs \cite{Cassani:2019vcl}. As in the maximally supersymmetric case, the form of the constrained compensator field \eqref{eq:LessSUSYCF} implies that
\begin{equation} \label{eq:GenParallFullLS}
		\gL_{\ff{J}_{A}} \ff{J}_{B} = X_{AB}{}^{C}\, \ff{J}_C \,,
\end{equation}
with $\ff{J}_A = \left( J_A,\, \Sigma_A \right)$.

\subsection{Truncation ansatz}
Using the $G$-structure $\cJ_A{}^M$ and $\rho$, we can now write down the consistent truncation ansatz to ${\cal N} < 16$ gauged supergravity. It is given as follows for the metric and vector fields
\begin{equation}
	\begin{split}
		g_{\mu\nu}(x,Y) &= \rho^{-2}(Y)\, G_{\mu\nu}(x) \,, \\
		A_\mu{}^M(x,Y) &= \cJ_A{}^M(Y)\, {\cal A}_\mu{}^{A}(x) \,, \\
		B_{\mu\,M}(x,Y) &= \Sigma_{A\,M}(Y)\, {\cal A}_{\mu}{}^{A}(x) \,.
	\end{split} \label{eq:TruncAnsatzLS}
\end{equation}
Here $G_{\mu\nu}$ is the metric and ${\cal A}_\mu{}^A$ the vector fields of the ${\cal N} \leq 16$ gauged supergravity. For the the scalar fields, we need to express the generalised metric $\gM_{MN} \in \EE/\SO{16}$ in terms of the $\Gst$-singlets $\cJ_A{}^M$ and $\rho$. While this is always possible since $\Gst \subset \SO{16}$, the resulting expression depends on the particular amount of supersymmetry kept.

For example, as we discuss in more detail in the next section, for ${\cal N}=12$ supersymmetry, we have $\Gst = \SU{2}$ such that the $\cJ_A{}^M$ correspond to the generators of $\mathrm{E}_{7(-5)} \subset \EE$ which commutes with $\SU{2} \subset \EE$. Let us further decompose $\mathrm{E}_{7(-5)}$ under its maximal compact subgroup $\mathrm{E}_{7(-5)} \subset \SO{12} \times \SO{3}$, so that
\begin{equation}
	\begin{split}
		\mbf{133} &\rightarrow \mbf{(66,1)} \oplus \mbf{(32,2)} \oplus \mbf{(1,3)} \,.
	\end{split}
\end{equation}
Accordingly we write for a generator of $\mathrm{E}_{7(-5)}$,
\begin{equation}
	t^A = \left( t^{IJ},\, t^{\alpha\,u},\, t^{uv} \right) \,,
\end{equation}
with $I, J = 1, \ldots, 12$ denoting the vector of $\SO{12}$, $\alpha, \beta = 1, \ldots, 32$ the spinor of $\SO{12}$ and $u, v = 1, 2$ the $\mbf{2}$ of $\SO{3}$. We can now write the generalised metric as
\begin{equation}
	\begin{split}
		\gM_{MN}(x,Y) &= \frac{1}{19}\, \rho^{-2}(Y)\, \cJ^A{}_M(Y)\, \cJ^B{}_N(Y) \left( 40\, M^{(66,1)}_{AB}(x) + 17\, M^{(32,2)}_{AB}(x) - 56\, M^{(1,3)}_{AB}(x) \right) \\
		& \quad + \frac{1}{19} \rho^{-2}(Y)\, \cJ^A{}_{MP}(Y)\, \cJ^B{}_{N}{}^P(Y) \left( 320\, M^{(66,1)}_{AB}(x) + 250\, M^{(32,2)}_{AB}(x) + 160\, M^{(1,3)}_{AB}(x) \right) \\
		& \quad - \eta_{MN} \,,
	\end{split}
\end{equation}
where $\cJ^A{}_{MN} = f_{MN}{}^P \cJ^A{}_P$ and we have defined the scalar-dependent inner products
\begin{equation} \label{eq:GenMetricN12}
	\begin{split}
		M^{(66,1)}_{AB}(x) &= b^{IJ}{}_A(x)\, b^{KL}{}_{B}(x)\, \delta_{IK}\,\delta_{JL} \,, \\
		M^{(32,2)}_{AB}(x) &= b^{\alpha\,u}{}_A(x)\, b^{\beta\,v}{}_{B}(x)\, \Omega_{\alpha\beta}\, \epsilon_{uv} \,, \\
		M^{(1,3)}_{AB}(x) &= b^{uv}{}_A(x)\, b^{wx}{}_B(x)\, \epsilon_{uw}\, \epsilon_{vx} \,.
	\end{split}
\end{equation}
Here $\Omega_{\alpha\beta}$ is the symplectic inner product on the $\mbf{32}$, $\epsilon_{uv}$ the $\SU{2}$-invariant alternating symbol and $b^{IJ}{}_A(x)$, $b^{\alpha\,u}{}_{A}(x)$ and $b^{uv}{}_A(x)$ are the scalar fields parameterising the coset space $\Msc = \frac{\mathrm{E}_{7(-5)}}{\SO{12} \times \SO{3}}$.

The condition \eqref{eq:GenParallFullLS} now ensures that the $Y$-dependence of all these objects factorises under the generalised Lie derivative. For example, we have that
\begin{equation}
	\begin{split}
		D_{\mu} g_{\nu\rho}(x,Y) &= \rho^{-2}(Y) \left( \partial_\mu G_{\nu\rho}(x) - {\cal A}_\mu{}^{A}(x)\, \theta_{A}\, G_{\nu\rho}(x) \right) \equiv \rho^{-2}(Y)\, {\cal D}_{\mu} G_{\nu\rho}(x) \,, \\
		\gL_{\ff{J}_\mu} \ff{J}_{\nu}(x,Y) &= \ff{J}_{C}(Y)\, X_{AB}{}^{C}\, {\cal A}_\mu{}^{A}(x)\, {\cal A}_\nu{}^{B}(x) \equiv \ff{J}_{A}(Y)\, \Leib{{\cal A}_{\mu}}{{\cal A}_{\nu}}^{A} \,,
	\end{split}
\end{equation}
and similarly for generalised metric, using its explicit expression such as \eqref{eq:GenMetricN12}. We thus find that the $\EE$ generalised Lie derivative reduces to the gauge-covariant derivative of the ${\cal N} < 16$ gauged supergravity, specified by the embedding tensor.

\section{Classifying ${\cal N} < 16$ gauged supergravities with higher-dimensional origin} \label{s:ClassLessSUSY}
We can now use the fact that we must have a structure group $\Gst \subset \EE$ that stabilises ${\cal N} < 16$ spinors, transforming in the $\mbf{16}$ of $\SO{16}$, to algebraically classify which 3-dimensional gauged supergravities have a higher-dimensional origin. There will, however, be further constraints, in order to have a consistent truncation which we will not address here. In particular, in order to ensure the consistency of the truncation, the intrinsic torsion must only contain singlets under $\Gst$ and these must be constant. This will impose differential conditions and restrict the allowed backgrounds, and the possible three-dimensional gaugings that can arise. For example, as discussed in \ref{s:CompGauge}, the largest possible compact gauging is $\SO{9}$, and in \ref{s:ProdTruncation} we cannot have consistent truncations on products of spheres where the full isometry group is gauged, even though these are clearly allowed from the perspective of the generalised structure group. 

For the purposes of this paper, we will limit ourselves to finding the possible matter contents that can arise from consistent truncations of type II/11-dimensional supergravity, and not discuss the gaugings. We will list the generalised structure groups $\Gst$ required for the truncations and the scalar manifolds that arise in separate sections, divided according to the various amounts of supersymmetry preserved by the truncation. Moreover, we will only consider Lie structure groups and there may be some additional possibilities coming from discrete $\Gst$.

Finally, we will also limit ourselves to the case where we have exactly ${\cal N}$ stabilised spinors leading to a gauged supergravity with ${\cal N}$ supersymmetries. We could, in principle, also consider truncations with more than ${\cal N}$ spinors stabilised by the structure group. However, in this case, the ${\cal N}$ gauged supergravity necessarily arises as a subtruncation of the larger supersymmetric gauged supergravity. The reason for this is identical to the one in higher dimensions and we refer the interested reader to \cite{Josse:2021put}.

Our results for which ${\cal N} > 4$ gauged supergravities can arise from consistent truncations are summarised in table \ref{tab:N>4Sum}. There we list the relevant scalar manifolds, the structure groups $\Gst$ and the maximum number of matter multiplets $p_{\rm max}$ that can be kept in a consistent truncation. Note that throughout this section we will denote the scalar manifold by $\Msc$, which should not be confused with the generalised metric used elsewhere in this paper.

We also study which ${\cal N}=4$ gauged supergravities can arise from consistent truncations. ${\cal N}=4$ supersymmetry in three dimensions imposes that the scalar manifold must be a quaternionic-K\"{a}hler space, or a product of two quaternionic-K\"{a}hler spaces, which need not be symmetric spaces. However, by the analogous argument as in five dimensions \cite{Josse:2021put}, a higher-dimensional origin via a consistent truncation imposes that these scalar manifolds must be symmetric spaces. We list in table \ref{tab:N=4Single} all the ${\cal N}=4$ gauged supergravities with scalar manifolds that are a single quaternionic-K\"{a}hler manifold that can arise from consistent truncations. In table \ref{tab:N=4Double}, we also list some of the possible ${\cal N}=4$ gauged supergravities whose scalar manifolds are the product of two quaternionic-K\"{a}hler manifolds that can arise from a consistent truncation. However, we will not be exhaustive in this last classification.

In the following subsections, we explain in detail the cases listed in tables \ref{tab:N>4Sum} -- \ref{tab:N=4Double} arise.

\begin{table}[H]
	\renewcommand{\arraystretch}{1.9}
	\setlength{\tabcolsep}{2em}
	\[
	\begin{array}{|c|c|c|c|}
		\hline
		\mathcal{N} & \Msc & \Gst & \text{Restrictions} \\ \hline
		12 & \frac{E_{7(-5)}}{\SO{12}\times\SU{2}} & \SU{2} & \\ \hline
		10 & \frac{E_{6(-14)}}{\SO{10}\times\U{1}} & \SU{3} & \\ \hline
		9 & \frac{F_{4(-20)}}{\SO{9}} & G_2 & \\ \hline
		8 & \frac{\SO{8,p}}{\SO{8} \times \SO{p}} & \Gst(p) = \SO{8-p} & p \leq 8 \\ \hline
		6 & \frac{\SU{4,p}}{\mathrm{S}[\U{4} \times \U{p}]} & \begin{array}{c} \Gst(0)=\SO{10}\\ \Gst(1) = \SU{5} \\ \Gst(1<p\leq4) = \SU{5-p} \times \U{1} \end{array} & p \leq 5 \\ \hline
		5 & \frac{\USp{4,2p}}{\USp{4} \times \USp{2p}} & \begin{array}{c} \Gst(0)=\SO{11}\\ \Gst(1) = \SU{2} \times G_2 \\ \Gst(2) = \SU{2} \end{array} & p \leq 2 \\ \hline
	\end{array}
	\]
	\caption{Summary of the possible truncations to ${\cal N} > 4$ gauged supergravities. $\Gst$ denotes the structure group, with the notation $\Gst(p)$ for $p$ matter multiplets, $\Msc$ the scalar manifold and the restrictions on the number of matter multiplets that can be kept in a truncation, where relevant.}\label{tab:N>4Sum}
\end{table}

\begin{table}[H]
	\renewcommand{\arraystretch}{1.9}
	\setlength{\tabcolsep}{2em}
	\[
	\begin{array}{|c|c|c|c|}
		\hline
		\Msc & \Gst & \text{Embedding} & \text{Restrictions} \\ \hline
		\frac{E_{6(6)}}{\SU{6}\times\SU{2}} & \U{1} & \eqref{BrU1} & \\ \hline
		\frac{\Ff}{\USp{6}\times\SU{2}} & \SU{2} &\eqref{BrU1} & \\ \hline
		\frac{\Gt}{\SO{4}} & \USp{6} &\eqref{BrU1} & \\ \hline
		\frac{\SO{4,p}}{\SO{4}\times\SO{p}} & \SO{8-p}\times\SU{2} &\eqref{BrSO8-p} & p \leq 6 \\ \hline 
		\frac{\SU{2,p}}{\mathrm{S}[\U{2}\times\U{p}]}	& \begin{array}{c}
			\Gst(1) = \SU{6} \\ 
			\Gst(1 < p \leq 5) = \SU{6-p}\times\U{1}
		\end{array} &\eqref{BrSU6-p} & p \leq 5 \\ \hline
		\frac{\USp{2,2p}}{\SU{2}\times\USp{2p}} & \begin{array}{c} \Gst(1) = \SO{7}\times\SU{2} \\
			\Gst(2) = \SU{2}\times\SU{2} \end{array} &\eqref{BrSU2xSU2} & p \leq 2 \\ \hline
	\end{array}
	\]
	\caption{Summary of the possible truncations to ${\cal N} = 4$ gauged supergravities, leading to a single quaternionic-K\"{a}hler manifold. $\Gst$ denotes the structure group, with the notation $\Gst(p)$ where there is a family of scalar manifolds, labelled by $p$ and $\Msc$ is the scalar manifold. We also list the embeddings of $\Gst \subset \SO{12}$ and give restrictions on the integer $p$, where relevant.}\label{tab:N=4Single}
\end{table}

\begin{table}[H]
	\renewcommand{\arraystretch}{1.9}
	\setlength{\tabcolsep}{2em}
	\[
	\begin{array}{|c|c|c|c|}
		\hline
		\Msc & \Gst & \text{Embedding} & \text{Restrictions} \\ \hline
		\frac{\Gt}{\SO{4}}\times\frac{\SU{2,1}}{\SUc{2}{1}} & \SU{3} & \eqref{BrSU3}& \\ \hline
		\frac{\Gt}{\SO{4}}\times\frac{\Gt}{\SO{4}} & \SO{3} & \eqref{BrSO3}& \\ \hline
		\frac{\SO{4,p}}{\SO{4}\times\SO{p}}\times\frac{\SO{4,q}}{\SO{4}\times\SO{q}} & \SO{4-p}\times\SO{4-q} & \eqref{BrSO4-p4-q}& p, q \leq 2 \\ \hline
		\frac{\SU{2,1}}{\SUc{2}{1}}\times\frac{\SO{4,p}}{\SO{4}\times\SO{p}} & \begin{array}{c}
			\Gst(p \leq 4) = \SO{6-p}\times\U{1} \\
			\Gst(p = 5, 6) = \U{1}
		\end{array} &\eqref{BrSO6-p} & p \leq 6 \\ \hline
		\frac{\SU{2,1}}{\SUc{2}{1}}\times\frac{\SU{2,p}}{\SUc{2}{p}} & \begin{array}{c}
			\Gst(1) = \SU{3}\times\SU{3} \\
			\Gst(p=2,3) = \mathrm{S}[\U{2}\times\U{4-p}]
		\end{array} & \eqref{BrSU3xSU3}& p \leq 3 \\ \hline
	\end{array}
	\]
	\caption{Summary of a subset of possible truncations to ${\cal N} = 4$ gauged supergravities, leading to a product of quaternionic-K\"{a}hler manifold. $\Gst$ denotes the structure group, with the notation $\Gst(p)$ where there is a family of scalar manifolds, labelled by $p$ and $\Msc$ is the scalar manifold. We also list the embedding of $\Gst$ in $\SO{12}$ and the restrictions on the matter that can be kept in a truncation, where relevant.}\label{tab:N=4Double}
\end{table}

\subsection{$\mathcal{N}=12,10,9$}
The structure groups for ${\cal N} = 12, 10, 9$ are easily identified. These require $\Gst \subset \SO{16}$ such that there are exactly ${\cal N}$ singlets in the $\mathbf{16}$. Moreover, the commutants of $\Gst$ inside $\EE$ and $\SO{16}$, which we denote as $\Com{\Gst}{\EE}$ and $\Com{\Gst}{\SO{16}}$, respectively, must give rise to the numerator and denominator of the scalar manifold of the supergravity, as for example summarised in \cite{deWit:2003fgi}. These considerations lead to the structure groups listed in the first three rows of table \ref{tab:N>4Sum}.

The $\Gst$ in table \ref{tab:N>4Sum} arise from the following decomposition of $\SO{16}$.
\begin{equation}
	\begin{split} \label{eq:N=12Gst}
		\SO{16} &\rightarrow \SO{12}\times\SU{2}\times\SU{2} \,, \\
		\SO{16}&\rightarrow \SO{10}\times\SU{4}\rightarrow\SO{10}\times\SU{3}\times\U{1} \,, \\
		\SO{16}&\rightarrow\SO{9}\times\SO{7}\rightarrow\SO{9}\times G_2 \,,
	\end{split}
\end{equation}
where in the ${\cal N}=12$ case, the $\Gst = \SU{2}$ corresponds to one of the two $\SU{2}$ factors in the first line of \eqref{eq:N=12Gst}

\subsection{$\mathcal{N}=8$}
The scalar manifold in this case is
\begin{equation}
	\mathcal{M}_p=\frac{\SO{8,p}}{\SO{8}\times\SO{p}} \,,
\end{equation}
where $p$ denotes the number of matter multiplets coupled to the theory. The associated structure groups are
\begin{equation}
	\begin{split} \label{eq:Gst8}
		\Gst(p)&=\SO{8-p} \,.
	\end{split}
\end{equation}
Note that for $p > 6$, the structure group would have to be discrete, or the identity. The structure groups \eqref{eq:Gst8} are embedded in $\SO{16}$ as
\begin{equation}
	\SO{16}\rightarrow\SO{8}\times\SO{8}\rightarrow\SO{8}\times\SO{p}\times\SO{8-p} \,,
\end{equation}
such that the $\mbf{16}$ decomposes as
\begin{equation}
	\mbf{16} \rightarrow \trep{8_v,1} \oplus \trep{1,8_s} \rightarrow \trep{8_v,1,1}\oplus\trep{1,1,8} \,.
\end{equation}
Note that here we crucially rely on the $\SO{8}$ triality, to replace the $\mbf{8_v}$ with $\mbf{8_s}$ in the decomposition of the $\mbf{16}$ under $\SO{8} \times \SO{8}$. While the $\mbf{8_v}$ contains further singlets under $\SO{8-p} \subset \SO{8}$, the $\mbf{8_s}$ does not. This ensures that the $\mbf{16}$ contains precisely 8 singlets under $\Gst = \SO{8-p}$.

The largest subgroup $\SO{8,p}$ of $\EE$ is $\SO{8,8}$, corresponding to $p = 8$. Therefore, we can keep at most $p = 8$ matter multiplets in the 3-dimensional ${\cal N}=8$ supergravity via a consistent truncation.

\subsection{$\mathcal{N}=6$}
The scalar manifolds and structure groups for 3-dimensional ${\cal N}=6$ gauged supergravity with $p$ matter multiplets are
\begin{equation}
	\begin{split}
		\mathcal{M}_p &=\frac{\SU{4,p}}{\SUc{4}{p}} \,,\\
		\Gst (0)&=\SO{10}\,, \\
		\Gst(1)&=\SU{5} \,, \\
		\Gst(p>1)&=\SU{5-p}\times\U{1} \,.
	\end{split}
\end{equation}
Note that there is no Lie structure groups that gives $p = 5$ and that $\SU{4,p} \nsubseteq \EE$ for $p > 5$. Therefore, we can keep at most $p=5$ matter mutliplets in a consistent truncation to ${\cal N}=6$ supergravity.
The $\Gst$ is embedded in $\SO{16}$ as
\begin{equation}
	\SO{16}\xrightarrow{p=0}\SU{4}\times\SO{10}\xrightarrow{p=1}\SU{4}\times\SU{5}\times\U{1}\xrightarrow{1<p\leq4}\SU{4}\times\mathrm{S}[\U{5-p}\times\U{p}]\times \U{1} \,.
\end{equation}

\subsection{$\mathcal{N}=5$}
The relevant scalar manifolds are
\begin{equation}
	\mathcal{M}_p=\frac{\USp{4,2p}}{\USp{4}\times\USp{2p}} \,.
\end{equation}
Firstly, notice that this scalar manifold can only arise for $p\leq2$ since $\EE$ does not contain a $\USp{4,6}$ subgroup. For $p=0,1$ we have $\Gst(0)=\SO{11}$ and $\Gst (1)=\SU{2}\times G_2$ embedded as
\begin{equation}
	\begin{split}
		\SO{16}&\xrightarrow{p=0} \USp{4}\times \SO{11} \,, \\
		\SO{11}&\xrightarrow{p=1}\SU{2}\times\SU{2}\times\SO{7}\rightarrow\SU{2}\times\SU{2}\times G_2 \,.
	\end{split}
\end{equation}
To get $p=2$ multiplets, we break 
\begin{equation}
	\SU{2}\times G_2\rightarrow \SU{2}\times\SU{2}\times\SU{2}\rightarrow \SU{2}\times\SU{2}_D \,,
\end{equation}
and take $\Gst(2)=\SU{2}_D$ where $\SU{2}_D$ is the diagonal subgroup of the first and third $\SU{2}$'s.

\subsection{$\mathcal{N}=4$}
Whereas the scalar manifolds for ${\cal N} > 4$ came in single families, corresponding to $p$ matter multiplets, the ${\cal N}=4$ supergravities are richer, with the scalar manifolds being any product of quaternionic-K\"{a}hler manifolds. This also means that the classification is more involved. We will first consider scalar manifolds consisting of only one quaternionic-K\"{a}hler manifold, and fully classify all possibilities. We will then study some examples of products of quaternionic-K\"{a}hler manifolds.

We begin with truncations leading to single quaternionic-K\"{a}hler manifolds. Since we want to have ${\cal N}=4$ supersymmetry, all the structure groups arise from the decomposition
\begin{equation}
	\begin{split}
		\SO{16}&\rightarrow\SO{4}\times\SO{12} \,, \\
		\mbf{16}&\rightarrow\trep{2,2,1}\oplus\trep{1,1,12} \,, \\
		\mbf{120}&\rightarrow \trep{3,1,1}\oplus\trep{1,3,1}\oplus\trep{1,1,66}\oplus\trep{2,2,12} \,, \\
		\mbf{128'}&\rightarrow\trep{2,1,32'}\oplus\trep{1,2,32} \,,	
	\end{split}
\end{equation}
with $\Gst\subset\SO{12}$. In particular, the truncation with no matter multiplets corresponds to structure group $\Gst(0)=\SO{12}$. Notice that the number of scalar fields in a truncation can be deduced by counting the singlets of $\Gst$ in the $\mbf{128'}$, more specifically
\begin{equation}\label{eq:singstrat}
	\#\text{scalars}=2\left(\{\# \text{singlets}\in\mbf{32}\}+\{\#\text{singlets}\in\mbf{32'}\}\right) \,.
\end{equation}
We find the relevant structure groups by looking for subgroups of $\SO{12}$ such that the right hand side of \eqref{eq:singstrat} matches the dimension of the various quaternionic-K\"ahler manifolds we are interested in. The $\Gst$ structure groups leading to single quaternionic-K\"{a}hler manifolds now come from the following decompositions.
\begin{enumerate}[start=1,label={\textbf{Br.\arabic*:}},ref=\textbf{Br.\arabic*}]
	\item \label{BrU1} The $\U{1}$ structure group commuting with $\En{6}\subset\EE$ (see the first row of table \ref{tab:N=4Single}) is obtained by breaking
	\begin{equation}
		\SO{12}\rightarrow \SU{2}\times\USp{6}\rightarrow\SU{2}\times\SU{3}\times\U{1} \,.
	\end{equation}
	From this we can also identify the following two structures in table \ref{tab:N=4Single}, in particular choosing the $\SU{2}$ that commutes with $\USp{6}$ gives us the $\Ff\subset\EE$ commutant, while $\USp{6}$ itself commutes with $\Gt$. 
	\item \label{BrSO8-p} The structure group for the first family are obtained by breaking 
	\begin{equation}
		\SO{12}\rightarrow \SU{2}\times \SU{2}\times\SO{8}\rightarrow \SU{2}\times\SU{2}\times\SO{8-p} \,,
	\end{equation}
	in analogy to the $\mathcal{N}=8$ case we must use the triality to exchange $\mbf{8_v}\leftrightarrow\mbf{8_s}$ before going to $\SO{8-p}$, this way we make sure to have exactly four stabilised spinors and thus get $\mathcal{N}=4$ supergravity. The vector representation of $\SO{12}$ branches according to 
	\begin{equation}
		\mbf{12}\rightarrow \trep{2,2,1}\oplus \trep{1,1,8} \,.
	\end{equation}
	\item \label{BrSU6-p} For the second family the relevant breaking is 
	\begin{equation}
		\begin{split}
			\SO{12} &\rightarrow \SU{6}\times\U{1}\quad (p=1) \,, \\
			\SU{6}  &\rightarrow \SU{p}\times\SU{6-p}\times\U{1}\quad (1<p\leq{6}) \,.
		\end{split}
	\end{equation}
	\item \label{BrSU2xSU2}To understand the final family we first note that the $p=1$ quaternionic-K\"ahler is the same as $\frac{\SO{4,1}}{\SO{4}}$ and therefore the structures also agree. The remaining case arises from 
	\begin{equation}
		\SO{12}\rightarrow\SU{2}\times\USp{6}\rightarrow\SU{2}\times\SU{2}\times\USp{4}.
	\end{equation}
\end{enumerate}

We now turn to the case of two quaternionic-K\"{a}hler manifolds.
\begin{enumerate}[start=5,label={\textbf{Br.\arabic*:}},ref=\textbf{Br.\arabic*}]
	\item \label{BrSU3}This arises by breaking 
	\begin{equation}
		\SO{12}\rightarrow\SU{2}\times\USp{6}\rightarrow \SU{2}\times\SU{3}\times\U{1} \,,
	\end{equation}
	so we break to an $\SU{3}$ subgroup of the $\USp{6}$ structure that commutes with $\Gt\times\SU{2}\subset\EE$.
	\item \label{BrSO3}The $\SO{3}$ structure that commutes with $\Gt\times\Gt$ comes from 
	\begin{equation}
		\USp{6}\rightarrow \SO{3}\times\SU{2} \,.
	\end{equation}
	\item \label{BrSO4-p4-q}The third structure comes from
	\begin{equation}
		\begin{split}
			\SO{12}\rightarrow \SO{4}\times\SO{8}\rightarrow\SO{4}\times\SO{4}\times\SO{4} \,,
		\end{split}
	\end{equation}
	we then use the identifications $\SO{4}\simeq\SU{2}\times\SU{2}$, $\SO{3}\simeq \SU{2}$
	\begin{equation}\label{eq:so3xso3}
		\so{12}\rightarrow\su{2}\oplus\su{2}\oplus\su{2}\oplus\su{2}\oplus\su{2}\oplus\su{2} \,.
	\end{equation}
	The $\SO{3}\times\SO{3}$ structure associated to $p,q=1$ comes from identifying the first $\SO{3}$ with the $\su{2}$ diagonal of the second and fourth factors in \eqref{eq:so3xso3} and the second with the diagonal of the fifth and sixth.
	\item \label{BrSO6-p}We break 
	\begin{equation}
		\SO{12}\rightarrow \SU{6}\times \U{1}\rightarrow\SU{4}\times\SU{2}\times\U{1}\times\U{1} \,,
	\end{equation}
	and then for $p=1$ we set $\Gst=\SO{5}\times\U{1}$ where we use $\SO{5}\simeq\USp{4}$ and take $\USp{4}\subset\SU{4}$, the appropriate $\U{1}$ factor is the one arising from $\SU{6}\rightarrow\SU{4}\times\SU{2}\times\U{1}$. The structures for $1<p<6$ are obtained by breaking $\SO{5}$ to the relevant subgroups while for $p=6$ we only keep the $\U{1}$. 
	\item \label{BrSU3xSU3}For $p=1$ we simply break the $\SU{6}$ structure associated to the single quaternionic K\"ahler $\frac{\SU{2,1}}{\mathrm{S}[\U{2}\times\U{1}]}$ as
	\begin{equation}
		\SU{6}\rightarrow\SU{3}\times\SU{3}\times\U{1} \,.
	\end{equation}
	Similarly, for $p=2$ consider 
	\begin{equation}
		\SU{6}\rightarrow\SU{2}\times\SU{2}\times\SU{2}\times\U{1} \,,
	\end{equation}
	and put $\Gst=\SU{2}\times\SU{2}\times\U{1}$, note that the choice of $\SU{2}$ facors depends on how the $\U{1}$ charges are assigned. Finally the $p=3$ case can be obtained from
	\begin{equation}
		\SU{6}\rightarrow\SU{3}\times\SU{2}\times\U{1} \,.
	\end{equation}
\end{enumerate}

\section{$S^5 \times \Sigma$ truncations of IIB} \label{s:Riemann}
Using the above formalism, we can construct consistent truncations of IIB supergravity on $S^5 \times \Sigma_2$, with $\Sigma_2$ a constant-curvature Riemann surface, preserving various amounts of supersymmetry. Because IIB supergravity on $S^5$ is generalised parallelisable (i.e. has $\Gst = \mathbf{1}$), the $S^5 \times \Sigma_2$ compactification has a $\U{1}$ structure from the Riemann surface. By identifying this $\U{1}_\Sigma$ on the Riemann surface with a $\U{1} \subset \SO{6}$ isometry of $S^5$, we can embed the $\Gst = \U{1}$ structure group in different ways into $\SO{16} \subset \EE$ and hence have different numbers of invariant spinors. This is the generalised geometry equivalent of performing a ``topological twist'' and results in three-dimensional gauged supergravities with different amounts of supersymmetry, see for example \cite{Cassani:2019vcl}. A particularly interesting example is where $\Sigma_2$ is a hyperbolic space and we can embed $\U{1}$ in such a way to obtain ${\cal N}=4$ gauged supergravity with an ${\cal N}=(2,2)$ AdS$_3$ vacuum, corresponding to the IIB AdS$_3$ vacuum of \cite{Maldacena:2000mw}.

\subsection{Matter content}
Here we will focus on this ${\cal N}=4$ case and describe the scalar manifold and gauging of the 3-dimensional gauged supergravity that arises. However, it is a straightforward exercise to compute the truncation obtained by identifying the $\U{1}_\Sigma$ with $\U{1} \subset \SO{6}$ differently, for example to find the truncations around the vacua of \cite{Couzens:2021tnv}. The ${\cal N}=4$ gauged supergravity is obtained by identifying $\U{1}_\Sigma$ with $\U{1}_D \subset \SO{6}$ whose commutant inside $\SO{6}$ is
\begin{equation}
	\Com{\U{1}_D}{\SO{6}} = \SU{2} \times \U{1} \times \U{1} \,.
\end{equation}
This $\U{1}$ can be identified with the following branching
\begin{equation} \label{eq:U1DSU4}
	\SU{4} \rightarrow \SU{2} \times \SU{2} \times \U{1} \rightarrow \SU{2} \times \U{1}_D \times \U{1} \,,
\end{equation}
where the $\U{1}$ we are interested in is $\U{1}_D \subset \SU{2}$.\footnote{The subscript $D$ on $\U{1}_D$ refers to the fact that we can equivalently describe this $\U{1}_D$ as the diagonal $\U{1}$ of the Cartan torus of $\SU{4}$.} Under this branching, the $\mbf{4}$ of $\SU{4}$ decomposes as
\begin{equation} \label{eq:SpinorDecomp}
	\mbf{4} \rightarrow \mbf{2}_1^0 \oplus \mbf{1}_{-1}^{1} \oplus \mbf{1}_{-1}^{-1} \,,
\end{equation}
with the superscript representing the $\U{1}_D$ charge and the subscript the charge under the other $\U{1}$.

To obtain an ${\cal N}=4$ supergravity, we take a diagonal of the $\U{1}_D \subset \SO{6}$ with $\U{1}_\Sigma$ of the Riemann surface. To make this explicit, consider the decomposition
\begin{equation}
	\begin{split}
		\SO{16} &\rightarrow \USp{8} \times \SU{2}_\Sigma \rightarrow \SU{4} \times \U{1} \times \SU{2}_\Sigma \\
		&\rightarrow \SU{2} \times \U{1}_D \times \U{1} \times \U{1} \times \SU{2}_\Sigma \\
		&\rightarrow \SU{2} \times \U{1}_D \times \U{1} \times \U{1} \times \U{1}_\Sigma \,,
	\end{split}
\end{equation}
such that the $\mbf{16}$ spinors of $\SO{16}$ branch as
\begin{equation}
	\begin{split}
		\mbf{16} &\rightarrow \mbf{(8,2)} \rightarrow \mbf{(4,2)}_1 \oplus \mbf{(\ov{4},2)}_{-1} \\
		&\rightarrow \mbf{(2,2)}_{1,1}^0 \oplus \mbf{(1,2)}_{1,-1}^1 \oplus \mbf{(1,2)}_{1,-1}^{-1} \oplus \mbf{(2,2)}^0_{-1,-1} \oplus \mbf{(1,2)}^{-1}_{-1,-1} \oplus \mbf{(1,2)}_{-1,1}^1 \\
		&\rightarrow \mbf{2}_{1,1}^{0,1} \oplus \mbf{2}_{1,1}^{0,-1} \oplus \mbf{1}_{1,-1}^{1,1} \oplus \mbf{1}_{1,-1}^{1,-1} \oplus \mbf{1}_{1,-1}^{-1,1} \oplus \mbf{1}_{1,-1}^{-1,-1} \\
		& \quad \oplus \mbf{2}^{0,1}_{-1,-1} \oplus \mbf{2}^{0,-1}_{-1,-1} \oplus \mbf{1}^{-1,1}_{-1,-1} \oplus \mbf{1}^{-1,-1}_{-1,-1} \oplus \mbf{1}_{-1,1}^{1,1} \oplus \mbf{1}_{-1,1}^{1,-1} \,,
	\end{split}
\end{equation}
where the first superscript is the $\U{1}_D$ charge, $q_D$, the second superscript is the charge $q_\Sigma$ under $\U{1}_\Sigma$, i.e. the Riemann surface holonomy group, and the subscripts refer to the charges under the other two $\U{1}$ subgroups. We now see that by taking $\U{1}_S$ as the diagonal of $\U{1}_D$ and $\U{1}_\Sigma$, with the charges
\begin{equation} \label{eq:qS}
	q_S = q_D + q_\Sigma \,,
\end{equation}
we have exactly 4 invariant spinors under $\U{1}_S$, so that the truncation will yield an ${\cal N}=4$ supergravity.

To compute the scalar manifold of the 3-dimensional supergravity, we need to determine the commutant of this $\U{1}_S$ inside $\EE$ and $\SO{16}$. To do this, we consider the branching
\begin{equation}
	\EE \rightarrow \En{6} \times \SL{3} \,,
\end{equation}
and further break
\begin{equation}
	\begin{split}
		\En{6} &\rightarrow \SL{6} \times \SL{2} \,, \\
		\SL{3} &\rightarrow \SL{2}_\Sigma \times \mathbb{R}^+ \,.
	\end{split}
\end{equation}
The $\U{1}_D$ is embedded as $\U{1}_D \subset \SO{6} \subset \SL{6}$, while $\U{1}_\Sigma \subset \SL{2}_\Sigma$. Using \eqref{eq:U1DSU4}, \eqref{eq:SpinorDecomp} and \eqref{eq:qS}, we find the following commutant of $\U{1}_D \subset \SL{6}$
\begin{equation}
	\Com{\U{1}_D}{\SL{6}} = \U{1}_D \times \SO{3,1} \times \SO{2,1} \times \mathbb{R}^+ \,,
\end{equation}
and hence
\begin{equation}
	\Com{\U{1}_D}{\En{6}} = \U{1}_D \times \SO{5,3} \times \mathbb{R}^+ \,,
\end{equation}
which comes from the branching
\begin{equation}
	\En{6} \rightarrow \SO{5,5} \times \mathbb{R}^+ \rightarrow \SO{5,3} \times \U{1}_D \times \mathbb{R}^+ \,.
\end{equation}
By identifying $\U{1}_S$ as the diagonal of $\U{1}_D$ with $\U{1}_\Sigma$, we now find the commutant in $\EE$
\begin{equation}
	\Com{\U{1}_S}{\En{8}} = \SU{2,1} \times \SO{6,4} \times \U{1}_S \,.
\end{equation}
Similarly, the commutant inside $\SO{16}$ is given by
\begin{equation}
	\Com{\U{1}_S}{\SO{16}} = \SU{2} \times \U{1} \times \SO{6} \times \SO{4} \times \U{1}_S \,.
\end{equation}
Note that this $\U{1}_S$ also precisely corresponds to the breaking \eqref{BrSO6-p} with $p=6$.

Hence we find that our consistent truncation has 28 scalar fields, parameterising the coset space
\begin{equation}
	{\cal M}_{\rm scalar} = \frac{\Com{\U{1}_S}{\EE}}{\Com{\U{1}_S}{\SO{16}}} = \frac{\SU{2,1}}{\SUc{2}{1}} \times \frac{\SO{4,6}}{\SO{4} \times \SO{6}} \,.
\end{equation}
The 53 generators of the numerator are associated to globally well-defined generalised vector fields in the $\mathbf{248}$ of $\En{8}$. To identify these, let us decompose $\EE \rightarrow \En{6} \times \SL{3}$
\begin{equation}
	\mbf{248} \rightarrow \mbf{\left(78,1\right)} \oplus \mbf{(1,8)} \oplus \mbf{(27,\overline{3})} \oplus \mbf{(\overline{27},3)} \,.
\end{equation}
The 53 generators of $\SU{2,1} \times \SO{4,6}$ now transform in the following representations under $\En{6} \times \SL{3} \rightarrow \SO{5,3} \times \U{1}_D \times \SL{2}_\Sigma$ such that they form singlets of $\U{1}_S$:
\begin{equation} \label{eq:U1Surface}
	\begin{split}
		\mathbf{\left( {27}, \overline{3} \right)} &\longrightarrow \mathbf{\left( 8_v,\, 1 \right)}_0 \oplus \mathbf{\left( 1, 2 \right)}_2 \oplus \mathbf{\left(1,2\right)}_{-2} \oplus \mathbf{\left(1,1\right)}_0 \,, \\
		\mathbf{\left( \overline{27}, 3 \right)} &\longrightarrow \mathbf{\left( 8_v,\, 1 \right)}_0 \oplus \mathbf{\left( 1, 2 \right)}_2 \oplus \mathbf{\left(1,2\right)}_{-2} \oplus \mathbf{\left(1,1\right)}_0 \,, \\
		\mathbf{\left(1,8\right)} & \longrightarrow \mathbf{\left(1,3\right)}_0 \oplus \mathbf{\left(1,1\right)}_0 \,, \\
		\mathbf{\left(78,1\right)} &\longrightarrow \mathbf{\left( 28,1 \right)}_0 \oplus 2 \cdot \mathbf{\left( 1,1 \right)}_0 \,.
	\end{split}
\end{equation}
Note that from each of the $\mathbf{\left(1,2\right)}_{\pm 2} \subset \mbf{(27,\overline{3})}$, $\mathbf{\left(1,2\right)}_{\pm 2} \subset \mbf{(\overline{27},3)}$ and $\mathbf{\left(1,3\right)}_0 \subset \mbf{(1,8)}$ there is exactly one singlet of $\U{1}_S$. Using the $\En{6}$ generalised parallelisation of type IIB on $S^5$ \cite{Hohm:2014qga,Lee:2014mla}, we can explicitly construct these 53 generators and compute the truncation Ansatz, but we will not do so here.

\subsection{Gauging}
We can compute the gauging by utilising that IIB supergravity admits a consistent truncation on $S^5$ to an $\SO{6} \subset \En{6}$ gauged supergravity. There are three effects that we need to take into account to obtain the gauging in three dimensions.
\begin{enumerate}
	\item The $\SO{6}$ gauging gets enhanced due to the dimensional reduction from five to three dimensions. In the $\EE$ framework, this arises because the $\En{6}$ embedding tensor can couple the $\mbf{248}$ generators and vector fields of $\EE$ in new ways, beyond the couplings that would arise in the $\En{6}$ theory in five dimensions.
	\item The gauging in three dimensions obtained this way gets broken because we only want to couple to the ${\cal N}=4$ sector, i.e. to the $\U{1}_S$ singlets that we keep in the consistent truncation.
	\item A non-trivial fibration of $S^5$ over the Riemann surface $\Sigma$ induces additional 3-dimensional embedding tensor components, which we need to compute.
\end{enumerate}

Let us now determine each of these effects in turn.

We begin by studying how the $\SO{6}$ gauging of the five-dimensional theory enhances upon dimensional reduction to three dimensions. The $\SO{6}$ gauging in five dimensions is due to an embedding tensor component in the $\mbf{351}$ of $\En{6}$. To understand the three-dimensional gauging that is induced, we decompose $\EE \rightarrow \En{6} \times \SL{3}$, upon which we have
\begin{equation}
	\begin{split} \label{eq:E8ToE6}
		\mbf{248} &\rightarrow \mbf{\left(27,\overline{3}\right)} \oplus \mbf{\left(\overline{27},3\right)} \oplus \mbf{\left(78,1\right)} \oplus \mbf{\left(1,8\right)} \,, \\
		\mbf{3875} &\rightarrow \mbf{\left(351,\overline{3}\right)} \oplus \mbf{\left(\overline{351},3\right)} \oplus \mbf{\left(27,\overline{6}\right)} \oplus \mbf{\left(\overline{27},6\right)} \oplus \mbf{\left(27,\overline{3}\right)} \oplus \mbf{\left(\overline{27},3\right)} \\
		& \quad \oplus \mbf{\left(78,8\right)} \oplus \mbf{\left(650,1\right)} \oplus \mbf{\left(1,8\right)} \oplus \mbf{\left(1,1\right)} \,.
	\end{split}
\end{equation}
The $\SO{6}$ gauging corresponds to a particular element of the $\mbf{\left(351,\overline{3}\right)}$. Let us write $M, N = 1, \ldots, 248$ for the fundamental of $\EE$, $A, B = 1, \ldots, 27$ for the fundamental representation of $\En{6}$ and $i, j = (\alpha, 3)$ for the fundamental of $\SL{3}$ with $\alpha, \beta = 1, 2$. Then the $\SO{6}$ gauging corresponds to the embedding tensor components
\begin{equation}
	\left(X_{\SO{6}}\right)_{AB}{}^{C,i} = \Theta_{AB}{}^C\, \delta^i_3 \,,
\end{equation}
with $\Theta_{AB}{}^C$ the $\En{6}$ embedding tensor corresponding to the $\SO{6}$ gauging.

This $\SO{6}$ embedding tensor now couples in the $\EE$ theory as
\begin{equation} \label{eq:SO6E8}
	\begin{split}
		A_\mu{}^M\, X_{MN}\, t^N &\sim A_\mu{}^A{}_i\, X_{AB}{}^{C,i}\, t^B{}_C + A_\mu{}^{B}{}_C\, X_{AB}{}^C\, t^A{}_i + A_{\mu\,C}{}^i\, X_{AB}{}^{C,j} t_D{}^k\, d^{ABD}\, \epsilon_{ijk} \\
		&= A_\mu{}^A{}_3\, \Theta_{AB}{}^{C}\, t^B{}_C + A_\mu{}^{B}{}_C\, \Theta_{AB}{}^{C}\, t^A{}_3 + A_{\mu\,A}{}^\alpha\, \Theta^{AB}\, t_B{}^\beta\, \epsilon_{\alpha\beta} \,,
	\end{split}
\end{equation}
where we have used the $\En{6}$ invariant to define $\Theta^{AB} = \Theta_{CD}{}^A\, d^{BCD}$. In the first term of \eqref{eq:SO6E8}, we recognise the coupling between the $\mbf{27}$ vector fields of five-dimensional gauged supergravity and the $\En{6}$ generators that we already had in the $\En{6}$ theory. Therefore, the first term reproduces the $\SO{6}$ gauging. On the other hand, the second and third terms are new, since they involve vector fields and generators outside $\En{6}$ that do not exist in the five-dimensional gauged supergravity. One can easily check that these extra terms enhance the $\SO{6}$ gauging by sets of commuting generators tranforming in the $\mbf{6}$ of $\SO{6}$. This would be the gauging that arises by reducing the five-dimensional $\SO{6}$ gauged supergravity on $T^2$. 

However, the gauging \eqref{eq:SO6E8} is broken by going to the ${\cal N}=4$ gauged supergravity with scalar coset space $\Msc = \frac{\SO{6,4}}{\SO{6} \times \SO{4}} \times \frac{\SU{2,1}}{\SUc{2}{1}}$. This is because the ${\cal N}=4$ gauging arises from the coupling to vector fields and generators that are singlets of $\U{1}_S$. As a result, this breaks the semi-simple $\SO{6}$ part of the gauging to its commutant with $\U{1}_D$, which is given by $\Com{\U{1}_D}{\SO{6}} = \SO{3} \times \U{1} \times \U{1}_D$. Moreover, of the commuting generators in the second and third term of \eqref{eq:SO6E8}, only those which are singlets of $\U{1}_S$ contribute.

The singlets of $\U{1}_S$ form representations $\SO{5,3} \times \mathbb{R}^+ = \Com{\U{1}_D}{\En{6}}$. Let us therefore decompose $\En{6} \rightarrow \SO{5,5} \times \mathbb{R}^+ \rightarrow \SO{5,3} \times \mathbb{R}^+ \times \U{1}_D$, such that
\begin{equation}
	\mbf{27} \rightarrow \mathbf{10}_2 \oplus \mbf{16}_{-1} \oplus \mbf{1}_{-4} \rightarrow \mbf{8_v}^0_2 \oplus \mbf{1}_2^2 \oplus \mbf{1}_2^{-2} \oplus \mbf{8_s}^{1}_{-1} \oplus \mbf{8_c}^{-1}_{-1} \oplus \mbf{1}^0_{-4} \,,
\end{equation}
where the superscript refers to the $\U{1}_D$ charge, while the subscript refers to the $\mathbb{R}^+$ charge. Accordingly, we write for a vector in the $\mbf{27}$ of $\En{6}$
\begin{equation}
	V^A = \left( V^{I},\, V^a,\, V^{\cal I},\, V^{\dot{\cal I}},\, V^z \right) \,,
\end{equation}
with $I = 1, \ldots, 8$ labelling the $\mbf{8_v}$, ${\cal I} = 1, \ldots, 8$ the $\mbf{8_s}$ and $\dot{\cal I} = 1, \ldots, 8$ the $\mbf{8_c}$ of $\SO{5,3}$, $a = 1, 2$ the $\U{1}_D$ doublet that comes from the $\mbf{10}$ of $\SO{5,5}$ and $z$ the $\SO{5,3} \times \U{1}_D$ singlet. By looking at $\U{1}_S$ invariants and knowing that the $\SO{6}$ gauging is compact and therefore does not couple to the $\mathbb{R}^+$ generator $t^z{}_z$, the gauging descending from \eqref{eq:SO6E8} in the ${\cal N}=4$ theory must be of the form 
\begin{equation} \label{eq:351N=4Group}
	\begin{split}
		A_\mu{}^M\, X_{MN}\, t^N &\sim A_\mu{}^I{}_3\, \Theta_{IJK}\, t^{JK} + A_\mu{}^I{}_3\, \Theta_{I\,ab}\, t^{ab} + A_\mu{}^z{}_3\, \Theta_{z\,IJ}\, t^{IJ} + A_\mu{}^z{}_3\, \Theta_{z\,ab}\, t^{ab} \\
		& \quad + A_\mu{}^{JK}\, \Theta_{IJK}\, t^I{}_3 + A_\mu{}^{ab}\, \Theta_{I\,ab}\, t^I{}_3 + A_\mu{}^{IJ}\, \theta_{z\,IJ} t^{z}{}_3 + A_\mu{}^{ab}\, \Theta_{z\,ab}\, t^z{}_3 \\
		& \quad + A_{\mu\,a}{}^\alpha\, \Theta^{ab}\, t_b{}^\beta\, \epsilon_{\alpha\beta} \,.
	\end{split}
\end{equation}

We can now evaluate \eqref{eq:351N=4Group} for the $\SO{6}$ gauging. To do this, let us decompose $\SO{5,3} \rightarrow \SO{3,1} \times \SL{2} \times \SL{2}$, so that we can identify the common subgroup of $\SO{5,3} \times \U{1}_D \times \mathbb{R}^+$ with $\SL{6} \times \SL{2}$, which is $\SO{3,1} \times \U{1}_D \times \SL{2} \times \SL{2} \times \mathbb{R}^+$. Table \ref{tab:SO31Dec} summarises how the $\mbf{27}$ of $\En{6}$ decomposes under these three subgroups, allowing us to match representations between $\SO{5,3} \times \U{1}_D \times \mathbb{R}^+$ and $\SL{6} \times \SL{2}$.

Immediately, we can see that the final term in \eqref{eq:351N=4Group} vanishes. This is because it only couples to vector fields and generators in the $\mbf{1}_2^{\pm 2}$ of $\SO{5,3} \times \U{1}_D \times \mathbb{R}^+$. However, from table \ref{tab:SO31Dec}, we see that these representations are only part of the $\mbf{(15,1)}$ of $\SL{6} \times \SL{2}$. However, for the $\SO{6}$ gauging, $\Theta^{AB}$ vanishes in the $\mbf{(15,1)} \otimes_{\rm antisym} \mbf{(15,1)}$ since there is no appropriate $\SO{6}$ invariants. Therefore, the final term of \eqref{eq:351N=4Group} gives no contribution for the case of $\SO{6}$.

Similarly, the first line gauges whatever $\SO{6}$ is broken to by $\U{1}_D$, thus the commutant $\Com{\U{1}_D}{\SO{6}}$, thus allowing us to also determine the second line of \eqref{eq:351N=4Group}. Let us therefore further break $\SO{5,3} \rightarrow \SO{5,3} \cap \left( \SO{6} \times \SL{2} \right) = \SO{3} \times \SO{2} \times \SL{2}$. Then we have the relevant decomposition
\begin{equation} \label{eq:SO53ToSO6SL2}
	\mbf{8_v} \rightarrow \mbf{(3,1)}_0 \oplus \mbf{(1,1)}_0 \oplus \mbf{(1,2)}_{\pm 1} \,.
\end{equation}
Let us denote by $u, v = 1, \ldots, 3$ the $\mbf{(3,1)}_0$, by $0$ the $\mbf{(1,1)}_0$. Similarly, let us denote by $y$ the $\SO{2}$ generator within the $\mbf{28}$ of $\SO{5,3}$. Then, \eqref{eq:351N=4Group} reduces to
\begin{equation} \label{eq:SO6N=4}
	\begin{split}
		A_\mu{}^M\, X_{MN}\, t^N &\sim A_\mu{}^u{}_3\,\epsilon_{uvw}\, t^{vw} + A_\mu{}^0{}_3\, \epsilon_{ab}\, t^{ab} + A_\mu{}^z{}_3\, t^y + A_{\mu}{}^{vw}\, \epsilon_{uvw}\, t^u{}_3 + A_{\mu}{}^{ab} \, \epsilon_{ab} \, t^0{}_3 + A_\mu{}^{y}\, t^z{}_3 \\
		&= A_\mu{}^u{}_3\,\epsilon_{uvw}\, t^{vw} + A_\mu{}^0{}_3\, t^0 + A_\mu{}^z{}_3\, t^y + A_{\mu}{}^{vw}\, \epsilon_{uvw}\, t^u{}_3 + A_{\mu}{}^{0} \, t^0{}_3 + A_\mu{}^{y}\, t^z{}_3 \,,
	\end{split}
\end{equation}
where $t^0 = \frac12 \epsilon_{ab} t^{ab}$ and $A_\mu{}^0 = \frac12 \epsilon_{ab} A_\mu{}^{ab}$.

\begin{table}[H]
	\renewcommand{\arraystretch}{1.6}
	\setlength{\tabcolsep}{2em}
	\[
	\begin{array}{|c|c|c|}
		\hline
		\SO{3,1} \times \SL{2} \times \SL{2} \times \U{1}_D \times \mathbb{R}^+ & \SO{5,3} \times \U{1}_D \times \mathbb{R}^+ & \SL{6} \times \SL{2} \\
		\hline
		\mbf{(1,2,2)}^0_2 & \mbf{8_v}^0_2 & \mbf{(\ov{6},2)} \\
		\mbf{(2 \otimes \ov{2}, 1, 1)}^0_2 & \mbf{8_v}^0_2 & \mbf{(15,1)} \\
		\mbf{(1,1,1)}^0_{-4} & \mbf{1}^0_{-4} & \mbf{(15,1)} \\
		\mbf{(1,1,1)}^2_2 & \mbf{1}^2_{2} & \mbf{(15,1)} \\
		\mbf{(1,1,1)}^{-2}_2 & \mbf{1}^{-2}_2 & \mbf{(15,1)} \\
		\mbf{(2,1,2)}^{-1}_{-1} & \mbf{8_c}_{-1}^{-1} & \mbf{(\ov{6},2)} \\
		\mbf{(\ov{2},2,1)}^{-1}_{-1} & \mbf{8_c}_{-1}^{-1} & \mbf{(15,1)} \\
		\mbf{(\ov{2},1,2)}^{1}_{-1} & \mbf{8_s}_{-1}^1 & \mbf{(\ov{6},2)} \\
		\mbf{(2,2,1)}^{1}_{-1} & \mbf{8_s}_{-1}^1 & \mbf{(15,1)} \\
		\hline
	\end{array}
	\]
	\caption{A dictionary relating representations of $\SL{6} \times \SL{2} \subset \En{6}$ and $\SO{5,3} \times \U{1}_D \times \mathbb{R}^+ \subset \En{6}$ via their common subgroup $\SO{3,1} \times \SL{2} \times \SL{2} \times \U{1}_D \times \mathbb{R}^+$. We focus on the representations appearing in the $\mbf{27}$ of $\En{6}$, which under $\SL{6} \times \SL{2}$ decomposes as $\mbf{27} \rightarrow \mbf{(15,1)} \oplus \mbf{(\ov{6},2)}$, while under $\SO{5,3} \times \U{1}_D \times \mathbb{R}^+$ decomposes as $\mbf{27} \rightarrow \mbf{8_v}_{2}^0 \oplus \mbf{1}_{2}^{\pm 2} \oplus \mbf{8_c}_{-1}^{-1} \oplus \mbf{8_s}_{-1}^1 \oplus \mbf{1}_{-4}^0$. Under $\SO{3,1} \times \SL{2} \times \SL{2} \times \U{1}_D \times \mathbb{R}^+$, these representations further decompose into the irreducible representations listed in the first column. In the second and third columns, we list the origin of the $\SO{3,1} \times \SL{2} \times \SL{2} \times \U{1}_D \times \mathbb{R}^+$ representations with respect to $\SO{5,3} \times \U{1}_D \times \mathbb{R}^+$ and $\SL{6} \times \SL{2}$. In the first and second columns, the subscripts refer to the $\mathbb{R}^+$ charge, while the superscripts refer to the $\U{1}_D$ charge. We denote by $\mbf{2}$ and $\mbf{\ov{2}}$ the fundamental and complex conjugate representations of $\SO{3,1}$.}\label{tab:SO31Dec}
\end{table}

Finally, we need to take into account the fibration of $S^5$ over the Riemann surface $\Sigma$, which is required to have a supersymmetric AdS vacuum in the three-dimensional theory. The effect of the fibration in the generalised geometry is that the $\EE$ generalised tangent bundle is not given by a direct sum of $\En{6}$ generalised tangent bundles tensored with appropriate factors of $T\Sigma$, etc, see for example \cite{Cassani:2019vcl}. Rather, the $\EE$ generalised tangent bundle is twisted, i.e. it is given by an extension of the $\En{6}$ bundles by those on $\Sigma$. To be more precise, the $\En{8}$ generalised tangent bundle is twisted by the following element:
\begin{equation}
	e^\Upsilon \,, \qquad \Upsilon = v \otimes E_0 \,,
\end{equation}
where $v$ is the local connection 1-form on $\Sigma$ and $E_0$ is the generalised vector field in the $\mathbf{27}$ of $\En{6}$ corresponding to the $\U{1}_D \subset \SO{6}$ isometry that is identified with $\U{1}_\Sigma$.

As usual, the twist by $\Upsilon$ yields a new component of the $\EE$ embedding tensor, in this case proportional to the Riemann surface curvature $\kappa = dv$, with $d$ the exterior derivative on $\Sigma$. Instead of computing the contribution of $\Upsilon$ to the embedding tensor directly, we can use group theory to fully determine it. We begin by nothing that the generalised vector field generating the $\U{1}_D \subset \SO{6}$ is the element of the $\mbf{27}$ of $\En{6}$ that transforms in $\mathbf{8_v}$ under $\SO{5,3} \subset \En{6}$. Explicitly, $E_0$ is the $\mbf{(1,1)}_0$ component of the $\mbf{8_v}$ after breaking $\SO{5,3} \cap \left( \SO{6} \times \SL{2} \right) = \SO{3} \times \SO{2} \times \SL{2}$, see \eqref{eq:SO53ToSO6SL2}. Inside $\EE$, $E_0 \in \mbf{(27,2)}_{2} \subset \mbf{(27,\overline{3})}$ with respect to $\En{6} \times \SL{2}_\Sigma \times \mathbb{R}^+ \subset \En{6} \times \SL{3}$.

Moreover, since $v$ is a 1-form on $\Sigma$, it should be thought of as transforming in the $\mbf{2}_{-3} \subset \mbf{8}$ of $\SL{2}_\Sigma \times \mathbb{R}^+ \subset \SL{3} = \Com{\En{6}}{\EE}$. Because the Riemann surface derivatives are valued in $\mbf{(1,2)}_{-3} \subset \mbf{(1,8)}$ with respect to $\En{6} \times \SL{2}_\Sigma \times \mathbb{R}^+ \subset \En{6} \times \SL{3}$, its Riemann surface exterior derivative $\kappa$ will be valued in $\mbf{(1,1)}_{-6}$ under $\En{6} \times \SL{2}_\Sigma \times \mathbb{R}^+ \subset \En{6} \times \SL{3}$. On the other hand, as discussed above, $E_0$ transforms as the $\mbf{(27,1)}_2 \subset \mbf{(27,\overline{3})}$ of $\En{6} \times \SL{2}_\Sigma \times \mathbb{R}^+ \subset \En{6} \times \SL{3}$. Therefore the embedding tensor component sourced by $\partial \Upsilon \sim (dv) \otimes E_0$ must transform in the representation $\mbf{(27,1)}_{-4}$ of $\En{6} \times \SL{2}_\Sigma \times \mathbb{R}^+$. Comparing with \eqref{eq:E8ToE6}, upon decomposing $\EE \rightarrow \En{6} \times \SL{3} \rightarrow \En{6} \times \SL{2}_\Sigma \times \mathbb{R}^+$, we see there is only one such component in the $\mbf{3875}$ of $\EE$: the $\mbf{(27,\overline{6})}$ of $\En{6} \times \SL{3}$. Hence, we have that the embedding tensor component sourced by the non-trivial fibration of $S^5$ over $\Sigma$ corresponds to
\begin{equation}
	X_{\rm new} \in \mathbf{(27,\overline{6})} \subset \mathbf{3875} \,.
\end{equation}
Viewing this as the symmetric tensor product of two $\mbf{248}$ representations of $\EE$, we see that it must come from
\begin{equation}
	\mbf{(\overline{27},3)} \otimes_{\rm sym} \mbf{(\overline{27},3)} \supset \mbf{(27,\overline{6})} \,.
\end{equation}
Therefore, this embedding tensor component couples to vectors and generators of the ${\cal N}=4$ supergravity as follows
\begin{equation}
	A_\mu{}^M\, \left(X_{\rm new}\right)_{MN}\, t^N = \theta'_I \left( A_\mu{}^I{}_3\, t^z{}_3 + A_\mu{}^z{}_3\, t^I{}_3 \right) \,,
\end{equation}
where
\begin{equation}
	\theta'_I = g\,\kappa\, \delta_I^0 \,.
\end{equation}
Here the index $0$ again denotes the $\mbf{(1,1)}_0$ direction in the $\mbf{8_v}$ of $\SO{5,3}$ under the branching \eqref{eq:SO53ToSO6SL2} to $\SO{3} \times \SO{2} \times \SL{2}$, corresponding to the $\U{1}_D \subset \SO{6}$ isometry. Moreover, $g$ is a numerical factor that can be determined either by computing explicitly the embedding tensor by constructing the generalised $\U{1}$ structure \eqref{eq:U1Surface} and evaluating its generalised Lie derivative, or by studying the 3-dimensional supergravity and ensuring that it has an ${\cal N}=(2,2)$ AdS vacuum.

We can now put everything together to present the gauging of the 3-dimensional ${\cal N}=4$ supergravity obtained. We have
\begin{equation} \label{eq:SigmaGauging}
	\begin{split}
		A_\mu{}^M\, X_{MN}\, t^N &= A_\mu{}^u{}_3\,\epsilon_{uvw}\, t^{vw} + A_\mu{}^0{}_3\, t^{0} + A_\mu{}^z{}_3 \left( t^y + g\, \kappa\, t^0{}_3 \right) \\
		& \quad + A_{\mu}{}^{vw}\, \epsilon_{uvw}\, t^u{}_3 + A_{\mu}{}^0 \, t^0{}_3 + \left( A_\mu{}^{y} + g\, \kappa\, A_\mu{}^0{}_3 \right) t^z{}_3 \,.
	\end{split}
\end{equation}
From the $\EE$ commutation relations, we can easily determine the gauging of the ${\cal N}=4$ gauged supergravity. We find that the gauging is $\mathrm{ISO}(3) \times \U{1}^4$ gauging, where the $t^{uv}$ generate the $\SO{3}$, the $t^u{}_3$ generate the three translations in the adjoint of $\SO{3}$, and the $t^{[ab]}$, $t^0{}_3$, $t^y + g\, \kappa t^0{}_3$ and $t^z{}_3$ generate the $\U{1}^4$. We can also easily compute the embedding of the gauging within the isometries of the scalar coset space $\SO{6,4} \times \SU{2,1}$. To do so, let us first identify the $\SO{6,4}$ and $\SU{2,1}$ generators in terms of the $\SO{5,3}$ basis used in \eqref{eq:SigmaGauging}. We have
\begin{equation} \label{eq:so64Decomp}
	\begin{split}
		\mathfrak{so}(6,4) &= \left\{ t^{IJ},\, t^{I}{}_3,\, t_I{}^3,\, \frac13 t^z{}_z + \frac23 t^3{}_3 \right\} \,,
	\end{split}
\end{equation}
and upon decomposing $\SU{2,1} \rightarrow \SU{2} \times \U{1}$, we have
\begin{equation} \label{eq:su21Decomp}
	\begin{split}
		\mathfrak{su}(2,1) &\rightarrow \mathfrak{su}(2) \oplus \mathfrak{u}(1) \oplus \mbf{2}_3 \oplus \mbf{2}_{-3} \,, \\
		\mathfrak{su}(2) &= \left\{ \frac13 t^z{}_z - \frac13 t^3{}_3 ,\, t^z{}_3,\, t_z{}^3 \right\} \,, \\
		\mathfrak{u}(1) &= \left\{ - \frac34 t^{0} + \frac34 t^y \right\} \,, \\
		\mbf{2}_3 \oplus \mbf{2}_{-3} &= \left\{ t^a{}_\alpha ,\, t_a{}^\alpha \right\} \,.
	\end{split}
\end{equation}
Thus, we see that the gauging embeds as $\SO{3} \subset \SO{5} \subset \SO{6} \subset \SO{6,4}$ and the three translations are $t^u{}_3 \subset \mbf{8_v}$ of $\SO{5,3} \subset \SO{6,4}$. Similarly, the $\U{1}^4$ can be identified with appropriate $\U{1}$'s in $\SO{6,4}$ and $\SU{2,1}$ using \eqref{eq:so64Decomp} and \eqref{eq:su21Decomp}.\footnote{Recall that $I, J = 1, \ldots, 8$ denote the $\mbf{8_v}$ of $\SO{5,3}$, so that $t^{IJ}$ are the $\mbf{28}$ of $\SO{5,3}$ which come from the $\mbf{(78,1)}$ of $\En{6} \times \SL{3} \subset \EE$, $t^{I}{}_3$ are the $\mbf{8_v}$ of $\SO{5,3}$ coming from the $\mbf{(27,\ov{3})}$ of $\En{6} \times \SL{3} \subset \EE$ and $t_I{}^3$ the $\mbf{8_v}$ of $\SO{5,3}$ coming from the $\mbf{(\ov{27},3)}$ of $\En{6} \times \SL{3}$. On the other hand, $t^z{}_z$ is the $\mathbb{R}^+$ generator that commutes with $\SO{5,3} \times \U{1}_D \subset \En{6}$ and comes from the $\mbf{(78,1)}$ of $\En{6} \times \SL{3}$, while $t^3{}_3$ is the $\mathbb{R}^+_\Sigma$ generator that comes from the $\mbf{(1,8)}$ of $\En{6} \times \SL{3}$ upon breaking $\SL{3} \rightarrow \SL{2}_\Sigma \times \mathbb{R}^+_\Sigma$. Finally, $t^z{}_3$, $t_z{}^3$, $t^a{}_\alpha$ and $t_a{}^\alpha$ correspond to the $\mbf{1}_0$ and $\mbf{1}_{\pm 2}$ of $\SO{5,3} \times \U{1}_D$ coming from the $\mbf{(27,\ov{3})}$ and $\mbf{(\ov{27},3)}$ of $\En{6} \times \SL{3}$, respectively. Finally, $t^0$ and $t^y$ are the $\U{1}$ generators within $\Com{\U{1}_D}{\SO{6}}$.}

While we determined the gauging completely by group-theoretic means, it is also straightforward to explicitly construct the generalised vector fields of $\EE$ that are stabilised by $\U{1}_S$ using \eqref{eq:U1Surface} and computing their generalised Lie derivative to determine the embedding tensor with precise coefficients.

\section{Conclusions} \label{s:Conclusions}
In this paper, we showed how to construct consistent truncations of 10-/11-dimensional supergravity to 3-dimensional gauged supergravities with various amounts of supersymmetry. Key to this was the use of $\EE$ ExFT, and the construction of constrained compensator fields from the $G$-structure underlying the truncation so that the generalised Lie derivative can be defined. The $G$-structures then have to close into an algebra under the generalised Lie derivative, providing the equivalent of the ``constant singlet intrinsic torsion'' condition used in higher dimensions. In the maximal case, the resulting embedding tensor transforms in exactly the representations allowed by the linear constraint of 3-d gauged supergravity. We obtain different amounts of supersymmetry, depending on how many spinors are stabilised by the $G$-structure of the truncations.

Using the $\EE$ setup, we proved various general results about which 3-dimensional gauged supergravities can be uplifted by consistent truncations. In particular, the largest possible compact subgroup is $\SO{9}$, although this cannot be realised due to the linear constraint of 3-dimensional gauged supergravity. This rules out many gauged supergravities, with large gauge groups, including $\EE$, $\SO{8} \times \SO{8}$, etc, constructed in \cite{Nicolai:2001sv}. We also showed that we cannot define a consistent truncation on the 7- or 8-dimensional product of spheres, in which all the isometries are gauged. Moreover, we also analysed which less supersymmetric 3-dimensional supergravities can arise from consistent truncations, and derived upper bounds on the number of matter multiplets.

Finally, we also constructed several new classes of consistent truncations of IIA/IIB supergravity to 3 dimensions. In the maximal case, we constructed the consistent truncations of IIA/IIB on $S^7$, which are related by an outer automorphism of $\SO{8,8}$ and lead to two different $\SO{8} \ltimes T^{28}$ gauged supergravities. These were constructed by making use of two different $\SL{8}$ subgroups of $\EE$, upon which the analysis reduces to the $\SL{n}$ twist equations of \cite{Hohm:2014qga} with $n = 8$. These truncation ans\"{a}tze also capture consistent truncations of IIA/IIB on hyperboloids, leading to $\mathrm{CSO}(p,q,r) \ltimes T^{p,q,r}$ gaugings. We also constructed a consistent truncation of IIB supergravity on $S^5$ times a Riemann surface. This results in a ${\cal N}=4$ 3-dimensional supergravity with scalar coset space $\frac{\SO{6,4}}{\SO{6} \times \SO{4}} \times \frac{\SU{2,1}}{\SU{2} \times U{1}}$ and contains an ${\cal N}=(2,2)$ AdS$_3$ vacuumm.

Our work opens up the possibility to systematically construct consistent truncations to 3-dimensional gauged supergravities, leading to many possible future routes of investigation. For example, which 3-dimensional ${\cal N} < 16$ supergravities can be uplifted and what are their uplifts? Similarly, which of the ${\cal N}=(8,0)$ AdS$_3$ vacua constructed in \cite{Deger:2019tem} can be uplifted? Our results already prohibit a higher-dimensional origin for most of the gauged supergravities in \cite{Deger:2019tem} and it would be interesting to study the uplift of the remaining handful of possibilities. Moreover, the analysis presented here is a crucial first step to computing the full Kaluza-Klein spectrum around any vacua of the ${\cal N}=16$ theories that can be uplifted by generalising \cite{Malek:2019eaz,Malek:2020yue} and the 3-dimensional half-maximal version thereof \cite{Eloy:2020uix}. We leave these exciting questions for future work.

\section*{Acknowledgements}
The authors thank Camille Eloy, Achilleas Passias, Henning Samtleben and Dan Waldram for useful discussions. MG and EM are supported by the Deutsche Forschungsgemeinschaft (DFG, German Research Foundation) via the Emmy Noether program ``Exploring the landscape of string theory flux vacua using exceptional field theory'' (project number 426510644).
	
\appendix
	
\section{$\EE$ conventions}
Throughout we raise/lower $\EE$ indices using the Cartan-Killing metric defined in terms of the $\EE$ structure constants as
\begin{equation} \label{eq:CKMetric}
	\eta^{MN} = \frac{1}{60} f^{MK}{}_L f^{NL}{}_K \,.
\end{equation}
	
The $\EE$ ExFT makes use of various projectors of the tensor product of two $\mbf{248}$ representations
\begin{equation}
	\mbf{248} \otimes \mbf{248} = \mbf{1} \oplus \mbf{248} \oplus \mbf{3875} \oplus \mbf{27000} \oplus \mbf{30380} \,.
\end{equation}
The projectors needed in the $\EE$ ExFT are the projector onto the adjoint
\begin{equation} \label{eq:ProjAdj}
	\left(\mathbb{P}_{248}\right)^{M}{}_N{}^K{}_L = \frac{1}{60} f^{M}{}_{NP} f^{PK}{}_L \,,
\end{equation}
as well as
\begin{equation} \label{eq:Proj}
	\begin{split}
		\left(\mathbb{P}_1\right)_{MN}{}^{KL} &= \eta_{MN}\, \eta^{KL} \,, \\
		\left(\mathbb{P}_{3875}\right)_{MN}{}^{KL} &= \frac17 \delta^M_{(N} \delta^K_{L)} - \frac1{56} \eta^{MK} \eta_{NL} - \frac1{14} f^{P}{}_{N}{}^{(M} f_{PL}{}^{K)} \,.
	\end{split}
\end{equation}

\bibliographystyle{JHEP}
\bibliography{NewBib}
	
\end{document}